\PassOptionsToPackage{hyperfootnotes=false}{hyperref}

\documentclass[a4paper]{cas-sc}

\usepackage[authoryear]{natbib}
\usepackage{amsmath,amssymb}
\usepackage{booktabs,tabularx,array}
\usepackage{rotating}
\usepackage{graphicx}
\usepackage{xurl}
\usepackage{placeins}   
\usepackage{orcidlink}
\graphicspath{{figures/}}

\newcolumntype{P}[1]{>{\raggedright\arraybackslash}p{#1\linewidth}}
\newcolumntype{Q}[1]{>{\centering\arraybackslash}p{#1\linewidth}}

\AtBeginDocument{\def\doi#1{\url{https://doi.org/#1}}}

\newcommand{\orcidicon}[1]{\texorpdfstring{\,\orcidlink{#1}}{}}
\makeatletter
\AtBeginDocument{\let\cas@printemails\printemails
  \renewcommand{\printemails}{\begingroup\let\orcidicon\@gobble\cas@printemails\endgroup}}
\makeatother
\ExplSyntaxOn
\RenewDocumentCommand \printorcid { }
  {
    \seq_if_empty:NF \g_stm_orcid_seq
      {
        \group_begin:
          \tex_let:D \thefootnote \relax \footnotetext
            { \raggedright \textsc{orcid}(s):\c_space_token \seq_use:Nn \g_stm_orcid_seq { ;~ } }
        \group_end:
      }
  }
\ExplSyntaxOff

\ExplSyntaxOn
\AddToHook{env/keywords/after}{ \box_gset_wd:Nn \g_stm_key_box { 0pt } }
\ExplSyntaxOff

\makeatletter
\newcommand{\appsection}[2]{%
  \section*{Appendix #1. #2}%
  \phantomsection\def\@currentlabel{#1}%
  \renewcommand{\thesection}{#1}%
  \setcounter{subsection}{0}%
  \setcounter{table}{0}\renewcommand{\thetable}{#1.\arabic{table}}%
  \setcounter{figure}{0}\renewcommand{\thefigure}{#1.\arabic{figure}}%
  \setcounter{equation}{0}\renewcommand{\theequation}{#1.\arabic{equation}}}
\makeatother
\makeatletter
\ExplSyntaxOn
\NewDocumentCommand \settabletype { }
  {
    \def\@captype{table}
    \__reset_tbl:
    \dim_set:Nn \l_tbl_width_dim { \linewidth }
    \cs_set_eq:NN \@makecaption \__make_tbl_caption:nn
    \centering \sffamily \small
  }
\ExplSyntaxOff
\makeatother

\begin{document}
\let\WriteBookmarks\relax
\def\floatpagepagefraction{1}
\def\textpagefraction{.001}

\shorttitle{Personal preference or system role? Battery swapping among UK EV users}
\shortauthors{J. Parmar et al.}

\title[mode=title]{Personal preference or system role? Battery swapping among UK electric vehicle users}

\author[1]{Janak Parmar\orcidicon{0000-0001-9284-5198}}
\cormark[1]
\ead{j.m.parmar@bham.ac.uk}

\affiliation[1]{organization={School of Geography, Earth and Environmental Sciences, University of Birmingham},
            city={Birmingham},
            country={United Kingdom}}

\author[2]{Huseyin Ayan\orcidicon{0009-0006-2850-1714}}
\ead{huseyin.ayan@iuc.edu.tr}

\affiliation[2]{organization={Department of Civil Engineering, Faculty of Engineering, Istanbul University-Cerrahpa\c{s}a},
            city={Istanbul},
            country={T\"urkiye}}

\author[1]{Dilum Dissanayake\orcidicon{0000-0001-6166-5709}}
\ead{d.dissanayake@bham.ac.uk}

\author[3]{Margaret Bell\orcidicon{0000-0001-8822-8437}}
\ead{margaret.bell@ncl.ac.uk}

\affiliation[3]{organization={School of Engineering, Newcastle University},
            city={Newcastle upon Tyne},
            country={United Kingdom}}

\cortext[cor1]{Corresponding author}

\begin{abstract}
Battery swapping replaces a depleted EV battery with a charged one, reducing waiting time and enabling off-peak charging. Evidence on acceptance, however, remains concentrated in Asian markets, and most studies examine individuals' adoption intentions. This study asks whether current UK EV users judge swapping differently as a purchase option and as a technology for the transport system. Using Behavioural Reasoning Theory, an integrated choice and latent variable model, and latent class analysis, we analyse perceived usefulness, energy-system concerns, practical barriers, and satisfaction with current charging. Usefulness and energy-system concerns relate more strongly to expectations about the system. Battery compatibility and station coverage relate more strongly to personal preference, while spare battery costs relate to expectations about network viability. User groups differ in how they balance benefits, barriers, and satisfaction with existing charging. Policy should address driver barriers and network viability as distinct but connected priorities.
\end{abstract}



\maketitle

\section{Introduction}\label{sec:intro}

Reducing road-transport emissions requires not only electric vehicles (EVs), but also ways of charging them that fit drivers' needs. Charging time, vehicle range, and the reliability and availability of public chargers remain important concerns, particularly for households without home charging. In the United Kingdom, the Zero Emission Vehicle mandate targets 33\% of new car sales in 2026 and 80\% by 2030 \citep{DfT2025ZEV}. Battery electric vehicles accounted for 26.2\% of new registrations in the first nine months of 2026 \citep{SMMT2026}. A Department for Transport survey of 848 EV drivers found that public-charger users most often called for better reliability, and 79\% wanted more charge points \citep{BritainThinks2022}. Around two in five UK households do not have a driveway \citep{DAmico2026}. Battery swapping offers a different way to replenish energy: a depleted battery is removed and replaced with a charged one, so drivers spend less time waiting. The service is well established in China. By the end of 2024, the largest operator, NIO, reported 3,054 stations and more than 61 million swaps \citep{NIO2025}. In Europe, however, deployment remains limited; Germany had 19 operational stations at the end of 2024 \citep{Kegel2025}. Swapping also separates the time of battery charging from the time a vehicle arrives, allowing operators to charge when electricity demand is lower. Engineering studies have examined this flexibility for peak-load management and grid services \citep{Cui2023,Revankar2021,Sen2026,Vallera2021,Wu2022}. But potential benefits for the electricity system do not tell us whether drivers themselves would choose swapping.

Evidence about public acceptance still comes mainly from Asian markets. Many passenger-car studies ask prospective buyers or members of the public, rather than people who already own and charge an EV. Recent work has begun to distinguish reasons for adopting battery swapping from reasons against it \citep{AduGyamfi2024b,Chopdar2026,Murugan2024}, but usually examines only one outcome: a person's intention to adopt the service. Behavioural Reasoning Theory (BRT) treats reasons for and against an action as separate influences whose importance depends on the situation \citep{Westaby2005}. This distinction is relevant to swapping because the service is useful only when enough compatible vehicles and stations are available. EV markets also have network effects: the service becomes more useful as more users and providers join, and it may need to reach a minimum scale to survive \citep{Li2017,ZhouLi2018}. A driver may expect swapping to benefit the transport system without wanting to buy a compatible vehicle. Another may like the service personally but doubt that a reliable network will emerge. A study focused only on personal intention cannot tell whether a factor matters more for an individual's purchase decision or for expectations about the wider system.

We examine these two judgements using survey responses from current EV users in the United Kingdom. An integrated choice and latent variable model estimates perceived usefulness, perceived pressure on the electricity system, reasons against swapping, personal purchase preference and expected system role within one framework. Because the model estimates the survey indicators and outcome relationships together, it accounts for measurement error without first converting the indicators into factor scores. We then use latent class analysis to identify groups of respondents who combine perceived benefits, barriers and satisfaction with current charging in different ways. Finally, we compare the groups on personal purchase preference and expected system role.

The study extends existing research in four ways. It brings evidence from the United Kingdom and from people who already drive and charge an EV into a literature focused largely on China and prospective buyers. It also moves beyond a single personal outcome by testing whether perceived benefits are more closely related to expectations about the system, while practical barriers are more closely related to purchase preference. In addition, it brings perceived pressure on the electricity system into a behavioural model of battery-swapping acceptance, an issue often discussed in engineering research but rarely measured in user studies. Finally, it identifies groups of EV users who combine reasons for and against swapping differently, helping show which concerns matter to which users.

The next section reviews previous research and develops the hypotheses. Section~\ref{sec:data} describes the survey and measures, and Section~\ref{sec:method} sets out the modelling approach. Section~\ref{sec:results} reports the results, which are discussed in relation to earlier studies in Section~\ref{sec:discussion}. Section~\ref{sec:policy-conclusions} draws out the policy implications, conclusions, limitations and priorities for future research.

\section{Literature review and hypotheses}\label{sec:lit}
This section first reviews empirical evidence on user acceptance of battery swapping and then introduces Behavioural Reasoning Theory and innovation resistance theory, together with their applications in transport and energy. It concludes by explaining why reasons for and against swapping may relate differently to personal and system-level judgements, and sets out the hypotheses.
\subsection{User acceptance of battery swapping}\label{sec:lit-acceptance}

Research on battery swapping has concentrated on engineering and operations, with less attention to user acceptance. A review of 78 studies focuses on station locations, grid interaction and routing \citep{Sen2026}. Another review found that only 14 of 303 studies of battery-as-a-service considered users' views \citep{Kopackova2026}. Our search of Scopus and Web of Science in October 2026 identified 26 empirical studies of acceptance. Sixteen concerned private passenger cars, and sixteen were conducted in China. Table~\ref{tab:studies} presents ten studies selected to reflect differences in markets, vehicle types, research designs and outcome measures. Other studies are discussed below.

The evidence gives a mixed picture of demand. Early stated-preference research in Japan found that swapping stations could benefit society once EV buyers made up a sufficient share of new-vehicle buyers \citep{Ito2013}. A later study, however, found that people preferred plug-in charging when driving range and station coverage were comparable \citep{Ito2019}. Evidence from Western Europe remains limited: a German survey found that respondents were willing to pay more for an EV with a swappable battery \citep{Kegel2025}, while a study of commercial transport reported positive but uncertain views \citep{Noto2023}. Across the wider literature, adoption or use intentions are commonly associated with perceived usefulness, attitude, knowledge, convenience and social influence \citep{AduGyamfi2022c,AduGyamfi2022a,AduGyamfi2022b,Murugan2024,XuS2025,Yan2026,Zhang2025a,Zhang2025b}. Studies of two-wheelers report similar patterns. Indonesian motorcyclists preferred station swapping or rapid charging to plugging in at home or work \citep{Balijepalli2023}, while a Taiwanese usability study found high acceptance of the better-designed of two swapping stations \citep{HuangFH2020}. Perceived risk is a recurring concern \citep{AduGyamfi2024a,Chen2024,Lu2026,XuS2025}. Among Chinese EV users, trust and risk also help explain how social and technical factors relate to acceptance \citep{Liu2026}. Common worries include battery quality, maintenance and responsibility for damage \citep{Geng2025,Kegel2025,Yan2026}. Findings are not entirely consistent: perceived behavioural control is significant in some studies but not others \citep{AduGyamfi2022b,Tan2022,XuX2026}, and one Chinese study found that performance expectations lost significance after other perceptions were included \citep{Chen2024}. Several frequently cited studies also come from the same research team and include analyses of the same sample \citep{AduGyamfi2022c,AduGyamfi2022a,AduGyamfi2022b}, so they do not provide fully independent evidence.

Three practical barriers recur across this literature. First, manufacturers use different battery sizes and designs, making it difficult for one network to serve vehicles from different brands \citep{Sen2026,Vallera2021}. This problem is particularly prominent among Indian two-wheeler users \citep{Chopdar2026}. Second, operators must invest in stations and maintain a reserve of charged batteries \citep{Avci2015,Ito2019,Noel2016}. Third, a swapping service is less useful when stations are scarce, even if each exchange is quick \citep{HuangY2021,Zhuge2021}. European deployment also faces organisational and regulatory barriers and competition from existing charging services \citep{Kopackova2026}. These issues correspond to the four barriers considered here: station coverage, inventory cost, battery standardisation and uncertainty about battery availability. Most acceptance studies apply the Technology Acceptance Model (TAM), the Theory of Planned Behaviour (TPB), UTAUT or related frameworks through structural equation modelling, or use stated-choice and valuation methods \citep{AduGyamfi2022b,AduGyamfi2024b,Geng2025,Ito2019,Kegel2025,XuS2025}. A previous hybrid choice study considered electric micromobility rather than passenger cars \citep{Zhang2025a}. User studies also rarely ask whether people think the electricity system is under pressure, despite the attention this issue receives in engineering research. Finally, studies typically assess either the personal value of swapping \citep{LeeKim2024} or its value to the wider system \citep{Park2019,Tan2022}, rather than both judgements from the same respondents. Section~\ref{sec:lit-applications} returns to this gap.

\begin{sidewaystable}
\caption{Selected empirical studies of user acceptance of battery swapping}\label{tab:studies}
\footnotesize
\begin{tabularx}{\linewidth}{@{}P{0.11}P{0.06}P{0.075}P{0.13}P{0.12}P{0.13}X@{}}
\toprule
Study & Country & Vehicle & Sample & Design and framework & Outcome (level) & Main findings \\
\midrule
\citet{Ito2019} & Japan & Car & 2,408 licence holders planning to buy & Choice experiment; random-parameter logit & Own vehicle choice (personal) & Recharging EV preferred to switching EV even with 30-min charging; choice probability of switching EV below 0.1 even with stations at 75\% of fuel stations \\
\citet{AduGyamfi2022b} & China & Car & 405 members of the public & CB-SEM; TAM, TPB, perceived risk & Adoption intention (personal) & Subjective norm strongest (0.28), perceived risk next ($-$0.26); usefulness, attitude, knowledge positive; perceived behavioural control not significant \\
\citet{Tan2022} & China & Car & 872 Hangzhou households (11\% own a BEV) & Contingent valuation; TPB covariates & Household WTP in taxes for station construction (system) & Mean 400.52 CNY a year; income, policy attention, experience, knowledge and environmental belief positive \\
\citet{AduGyamfi2024b} & China & Car & 552 drivers & CB-SEM; TAM and innovation resistance & Adoption intention (personal) & Usage and value barriers negative; risk barrier not significant; enablers explain more variance than barriers \\
\citet{LeeKim2024} & South Korea & Car & 800 members of the public (2\% EV owners) & Contingent valuation & WTP for own use of a swapping service (personal) & Mean WTP KRW 107.4 thousand (USD 78.8); charging inconvenience raises WTP \\
\citet{Geng2025} & China & Car & 1,509 drivers at dealerships & Choice experiment; mixed logit & Own choice of swap, charge or petrol car (personal) & Swapping attracts drivers seeking lower purchase price and faster replenishment; concern over third-party battery maintenance \\
\citet{Kegel2025} & Germany & Car & 1,902 car owners and prospective buyers, incl. 309 BEV owners & Survey with embedded choice experiment & Own vehicle choice (personal) & Premium for a swappable over an integrated battery in every group, from about 22 to 56 euros a month; liability for battery damage the main concern \\
\citet{Zhang2025a} & China & E-bike, e-moped & 314 users and potential adopters & Choice experiment; hybrid choice model & Own choice of swappable vs rechargeable vehicle (personal) & Social norms and annual sales growth positive; risk perception, capital and running cost negative \\
\citet{Chopdar2026} & India & E-two-wheeler & 317 current users with swapping experience & PLS-SEM; BRT with formative reasons & Attitude; own use behaviour (personal) & Reasons against outweigh reasons for on use ($-$0.401 vs 0.199); interoperability the most heavily weighted reason against \\
\citet{Yan2026} & China & Taxi, ride-hailing car & 996 drivers who swapped in the previous three months & PLS-SEM; extended TAM & Continuance intention (personal) & Battery quality assessment strongest predictor; usefulness and ease of use positive \\
\bottomrule
\end{tabularx}

\vspace{3pt}
\begin{minipage}{\linewidth}\footnotesize \textit{Note:} These ten studies were selected from 26 studies found in Scopus and Web of Science on 16 October 2026. They cover different markets, vehicle types, research designs and outcome levels; the other studies are discussed in Section~\ref{sec:lit-acceptance}. WTP means willingness to pay; CB-SEM and PLS-SEM are covariance-based and partial least squares structural equation modelling; TAM is the Technology Acceptance Model; TPB is the Theory of Planned Behaviour; and BRT is Behavioural Reasoning Theory.\end{minipage}
\end{sidewaystable}

\subsection{Behavioural Reasoning Theory and innovation resistance}\label{sec:lit-brt}

Behavioural Reasoning Theory separates reasons for and against an action from broader influences such as attitude, social expectations and perceived control \citep{Westaby2005}. Specific reasons can affect intention directly or work through broader attitudes, and a reason for acting does not have to be the mirror image of a reason against it. For instance, limited charging infrastructure may discourage EV purchase even when a person has no equally strong positive reason for buying an EV \citep{Chatzidakis2013,Claudy2015}. The original BRT studies showed that these specific reasons explained intentions beyond broader motives, although their influence varied by outcome \citep{Westaby2005}. Later applications have often simplified the framework and focused mainly on attitude \citep{Claudy2013,Sahu2020}. Innovation resistance theory draws a related distinction between practical barriers, such as concerns about use, value and risk, and psychological barriers linked to habit, tradition or image \citep{RamSheth1989}. Resistance may involve delaying adoption, rejecting an innovation or opposing it. It can arise from concerns about the innovation itself or from satisfaction with the existing option \citep{Heidenreich2015,Kleijnen2009}. The two frameworks can be combined by treating these barriers as reasons against adoption \citep{Claudy2015}. In this study, the reasons-against items capture practical barriers to swapping, while satisfaction with current charging is measured separately because it reflects the appeal of the existing alternative.

The balance between positive and negative reasons is not the same across technologies. Reasons against were stronger for micro wind turbines, whereas reasons for were stronger for car sharing \citep{Claudy2015}. In research on household solar panels, reasons against affected intention directly, while reasons for operated through attitude \citep{Claudy2013}. We therefore do not assume that either type of reason is always more influential. Instead, we test whether their relative importance changes between personal purchase preference and expectations about the wider system. Studies have also operationalised reasons in different ways, from lists of specific concerns to reflective and formative constructs \citep{Chen2021,Chopdar2026,Claudy2013,Westaby2005}. \citet{Chopdar2026}, for example, treat reasons for and against swapping as formative higher-order constructs, whereas other studies use combined scores or latent variables in structural equation models.

\subsection{Applications in transport and energy}\label{sec:lit-applications}

Behavioural Reasoning Theory and innovation resistance theory have been applied to electric cars and two-wheelers, cycling, car sharing, autonomous vehicles and household energy technologies (Table~\ref{tab:brt} in Appendix~\ref{app:brt}). Lower costs, environmental benefits and convenience often encourage adoption; costs, weak infrastructure, range limits and safety concerns can discourage it \citep{Bhat2025,FatahUddin2024,NguyenPhuoc2023,Yadav2024}. The balance between these influences depends on the setting. Several Indian EV studies report stronger reasons for adoption \citep{Bhat2025,FatahUddin2024,Yadav2024}, whereas research on Vietnamese motorcyclists found that reasons against acted directly on switching intention while reasons for worked through broader motives \citep{NguyenPhuoc2023}. Barriers also feature prominently in studies of scooter use, cycling and resistance to EVs \citep{Chen2021,ClaudyPeterson2014,JuKim2022,Long2025}.

Two limitations in this literature are especially relevant here. First, many studies focus on people who might adopt an innovation rather than current EV users, a gap also noted by \citet{Bhat2025}. Second, most examine a person's own attitude, intention or behaviour. Even studies that compare adoption and resistance usually model two personal outcomes \citep{KohYuen2024}, while others compare technologies but still focus on a single outcome \citep{Garibbo2025}. In battery-swapping research, studies that assess wider system benefits do not also model the same respondents' personal choices and their reasons for and against swapping \citep{Park2019,Tan2022}. We therefore do not yet know whether the same benefit or barrier relates differently to personal preference and expectations about wider adoption.

\subsection{Personal and system-level outcomes: hypothesis development}\label{sec:hypotheses}

For technologies that depend on supporting infrastructure, a person's own preference can differ from their expectation of what will happen more broadly. Research on energy technologies distinguishes social and political acceptance, community acceptance and market acceptance \citep{Moesker2024,Wustenhagen2007}. Risk research similarly shows that personal and societal judgements can respond differently to the same information \citep{TylerCook1984}. In one US study, people preferred less vehicle automation for themselves than for other drivers \citep{Agarwal2024}. These findings caution against treating perceived benefits to the transport system as equivalent to a willingness to adopt a technology personally.

Research in energy and transport also suggests that personal adoption and wider support may have different drivers. Across the European Union, adoption of energy innovations was only weakly related to support for energy policies, and the two outcomes had different predictors \citep{Spandagos2022}. A meta-analysis found that general acceptance of energy technologies was more closely related to support than to purchase intention \citep{Milani2024}. Similarly, public support for bike-sharing can remain high even when relatively few people intend to cycle themselves \citep{Nikitas2018}. The relationship can also work in the other direction: expectations about future market shares shaped responses to information about zero-emission-vehicle policy \citep{Akiyama2025}, while expected sales growth influenced choices of swappable electric two-wheelers \citep{Zhang2025a}.

Personal preference and expectations about the system are connected, but they are not interchangeable. BRT suggests that reasons for and against an action may relate to these judgements in different ways. \citeauthor{Westaby2005}'s (\citeyear{Westaby2005}) original studies found that specific reasons added explanatory power beyond broader motives, with effects that varied by outcome. Later applications often show that reasons against act directly on intention, while reasons for work through attitude \citep{Claudy2013,Claudy2015,NguyenPhuoc2023}. If positive and negative reasons operate through different routes, their importance may also differ between personal and system-level judgements. Perceived usefulness captures the benefits people associate with swapping and may therefore be especially relevant to expectations about its wider role. Practical barriers such as limited station coverage and incompatible batteries, by contrast, are obstacles that drivers may face when deciding whether to buy a vehicle. They may consequently bear more directly on personal preference. This distinction reflects the difference between supporting an innovation and taking on the costs of adopting it \citep{Spandagos2022}.

The opposite pattern is also plausible. \citet{Milani2024} found that perceived benefits were more strongly associated with intention to buy an energy technology than with intention to support it. However, both outcomes concerned an individual's intention, whereas our system-level outcome asks what role respondents expect battery swapping to play in transport more generally. The earlier finding therefore does not settle which outcome should be more closely related to usefulness. We test the two relationships in the same model rather than infer a difference by comparing separate studies.

H1. Perceived usefulness is more strongly related to expected system role than to personal purchase preference.

H2. Reasons against battery swapping are more strongly related to personal purchase preference than to expected system role.

H3. The difference between the patterns in H1 and H2 is non-zero. That is, usefulness and barriers differ in which outcome they relate to, not merely in the direction of their relationships.

Perceived pressure on the electricity system is modelled separately because its indicators concern the electricity system, not the features of battery swapping. We do not specify a directional hypothesis for this construct; its relationships with both outcomes are examined in Section~\ref{sec:results}.

These hypotheses describe average relationships across respondents, but drivers may not all assess swapping in the same way. Research on innovation resistance distinguishes between delaying adoption, rejecting a technology and opposing it; these responses can reflect concerns about the innovation or satisfaction with the existing option \citep{Heidenreich2015,Kleijnen2009}. Latent class analysis has identified comparable groups among prospective buyers of zero-emission vehicles \citep{Long2025}. If users combine benefits and barriers differently, they may also differ in the gap between their personal preference and their expectations about the system. We therefore examine this question alongside the hypotheses.

RQ. Do current EV users form distinct groups based on how they combine reasons for and against battery swapping, and do these groups differ in personal purchase preference and expected system role in ways consistent with H1 and H2?

\section{Data and measures}\label{sec:data}
This section describes the survey and the sample, defines the two outcome variables and the explanatory constructs, and lists the covariates used in the models.
\subsection{Survey design and data collection}\label{sec:survey}

We collected data through an online survey of current EV users living in the United Kingdom. Respondents were recruited through an online panel, and the survey ran from June to July 2025. Following screening questions, the questionnaire covered four areas. The first concerned household vehicle ownership, reasons for acquiring an EV, vehicle range and travel patterns, including frequency of use, typical daily distance and trip purposes. The second asked about charging: where and how often respondents charged, charging duration and time of day, the battery level at which they usually recharged, and satisfaction with local public-charging provision. The third presented sixteen statements about battery swapping and the electricity system, rated from strongly disagree to strongly agree on a five-point scale; one was an attention check. The final section collected demographic information, including age, gender, education, household income and region. Research colleagues reviewed the questionnaire and suggested refinements before the survey went live.

Private-car battery swapping is not commercially available in the UK, so respondents were judging a service they had not used. To give them a common basis for assessment, the survey explained the process and its differences from plug-in charging using text and pictures. We restricted the sample to current EV users because they have first-hand experience of charging and can compare swapping with the arrangements they use today. The survey collected 795 responses; 726 were retained after response-quality checks.

\subsection{Sample characteristics}\label{sec:sample}

Respondents were generally experienced EV users who drove frequently and relied mainly on home charging. Their education and household income were also higher than those of the general population (Table~\ref{tab:sample}). The sample was close to evenly divided by gender, with 51.4\% women, and 78.4\% of respondents were aged 25 to 54. Around 72.1\% held a university degree, while 59.8\% reported a household income of at least \pounds{}60,000.

\begin{table}
\caption{Sample characteristics (N = 728)}\label{tab:sample}
\footnotesize
\setlength{\tabcolsep}{4pt}
\begin{tabularx}{\linewidth}{@{}Xrr@{\hspace{14pt}}Xrr@{}}
\toprule
Characteristic and category & $n$ & \% & Characteristic and category & $n$ & \% \\
\midrule
\multicolumn{3}{@{}l}{\textit{Gender}} & \multicolumn{3}{@{}l}{\textit{Housing and parking}} \\
\quad Female & 375 & 51.5 & \quad House with dedicated parking & 611 & 83.9 \\
\quad Male & 353 & 48.5 & \quad House without dedicated parking & 60 & 8.2 \\
\multicolumn{3}{@{}l}{\textit{Age}} & \quad Flat with off-road parking & 47 & 6.5 \\
\quad 18 to 24 & 36 & 4.9 & \quad Flat without parking & 8 & 1.1 \\
\quad 25 to 34 & 186 & 25.5 & \quad Other & 2 & 0.3 \\
\quad 35 to 44 & 217 & 29.8 & \multicolumn{3}{@{}l}{\textit{Charging locations used (multiple)}} \\
\quad 45 to 54 & 168 & 23.1 & \quad Home & 617 & 84.8 \\
\quad 55 to 64 & 90 & 12.4 & \quad Workplace & 171 & 23.5 \\
\quad 65 or over & 31 & 4.3 & \quad Public charge point & 297 & 40.8 \\
\multicolumn{3}{@{}l}{\textit{Highest education}} & \multicolumn{3}{@{}l}{\textit{Public charging frequency}} \\
\quad Secondary school or below & 131 & 18.0 & \quad Daily & 30 & 4.1 \\
\quad Professional qualification, no degree & 70 & 9.6 & \quad Weekly & 176 & 24.2 \\
\quad Undergraduate degree & 312 & 42.9 & \quad Monthly & 130 & 17.9 \\
\quad Postgraduate degree & 213 & 29.3 & \quad Infrequently & 392 & 53.8 \\
\quad Prefer not to say & 2 & 0.3 & \multicolumn{3}{@{}l}{\textit{Satisfaction with public charging availability}} \\
\multicolumn{3}{@{}l}{\textit{Household income}} & \quad 1, fully dissatisfied & 52 & 7.1 \\
\quad Less than \pounds{}20,000 & 26 & 3.6 & \quad 2 & 171 & 23.5 \\
\quad \pounds{}20,000 to \pounds{}39,999 & 109 & 15.0 & \quad 3 & 217 & 29.8 \\
\quad \pounds{}40,000 to \pounds{}59,999 & 135 & 18.5 & \quad 4 & 194 & 26.6 \\
\quad \pounds{}60,000 to \pounds{}79,999 & 149 & 20.5 & \quad 5, fully satisfied & 94 & 12.9 \\
\quad \pounds{}80,000 to \pounds{}99,999 & 128 & 17.6 & \multicolumn{3}{@{}l}{\textit{EV use frequency}} \\
\quad \pounds{}100,000 or more & 158 & 21.7 & \quad 5 days a week or more & 465 & 63.9 \\
\quad Prefer not to say & 23 & 3.2 & \quad 3 to 4 days a week & 186 & 25.5 \\
\multicolumn{3}{@{}l}{\textit{EV experience}} & \quad 1 to 2 days a week & 66 & 9.1 \\
\quad Less than 6 months & 37 & 5.1 & \quad 3 times a month or less & 11 & 1.5 \\
\quad 6 months to 1 year & 137 & 18.8 & \multicolumn{3}{@{}l}{\textit{Typical daily distance}} \\
\quad 1 to 3 years & 375 & 51.5 & \quad Less than 10 km & 150 & 20.6 \\
\quad More than 3 years & 179 & 24.6 & \quad 10 to 30 km & 332 & 45.6 \\
\multicolumn{3}{@{}l}{\textit{Electric range of main EV}} & \quad 30 to 50 km & 163 & 22.4 \\
\quad Less than 200 km & 158 & 21.7 & \quad More than 50 km & 83 & 11.4 \\
\quad 200 to 300 km & 231 & 31.7 & \multicolumn{3}{@{}l}{\textit{EV used for commuting}} \\
\quad 300 to 400 km & 208 & 28.6 & \quad Yes & 573 & 78.7 \\
\quad 400 to 500 km & 95 & 13.0 & \multicolumn{3}{@{}l}{\textit{Reasons for acquiring an EV (multiple)}} \\
\quad More than 500 km & 36 & 4.9 & \quad Cost savings & 581 & 79.8 \\
 & &  & \quad Environmental concerns & 439 & 60.3 \\
 & &  & \quad Technological interest & 292 & 40.1 \\
 & &  & \quad Government incentives & 118 & 16.2 \\
 & &  & \quad Other & 42 & 5.8 \\
\bottomrule
\end{tabularx}
\end{table}

EV use was both established and frequent. Around three quarters had driven an EV for more than a year, nearly two thirds used it at least five days a week, and 78.7\% used it for commuting. Most charged at home: 84.8\% did so, and 83.9\% lived in a house with dedicated parking. Public charging was a regular part of life for a smaller group. In total, 28.3\% used public chargers at least weekly, while 53.8\% used them only occasionally. Views on local public-charging availability were mixed: 39.5\% were satisfied and 30.6\% dissatisfied. Respondents came from all eleven regions of Great Britain, including 26 from Wales, 105 from the South East, 97 from Scotland and 77 from London.

The online panel was neither designed to represent all UK drivers nor weighted to do so. Only 9.3\% of respondents lacked off-street parking, compared with around two in five UK households without a driveway \citep{DAmico2026}. This partly reflects the profile of EV owners, who are more likely to have access to home charging. The findings should therefore be read as describing experienced EV users in this panel, most of whom charge at home, not UK drivers as a whole. People without home charging are particularly relevant because swapping could provide an alternative to public charging. The model accounts separately for the 111 respondents who did not charge at home (Section~\ref{sec:covariates}).

\subsection{Outcome variables}\label{sec:outcomes}

The outcomes are based on three questions from the survey section on the potential for battery-swapping adoption. Respondents rated each statement on a five-point scale from strongly disagree to strongly agree. The questions refer either to the respondent's own purchase or to the wider transport system; that difference defines the two outcomes.

\begin{itemize}
\item Personal purchase preference is measured by R1: ``I prefer to buy electric vehicles if battery swapping facility is available in general.'' We treat the response as a single ordered outcome.
\item Expected system role is measured by two statements: R2, ``Battery swapping potential would increase the electric vehicle uptake,'' and R3, ``Battery swapping could play a crucial role in shaping the future of transportation.'' Both ask about the vehicle market and transport system rather than the respondent's own purchase. Together, they measure a latent outcome representing the expected role of swapping in the wider system.
\end{itemize}

The three questions followed the attitudinal statements. Agreement rose as the question moved from personal purchase to the wider system: 45.5\% agreed with R1, 58.4\% with R2 and 71.0\% with R3. The mean responses showed the same pattern. For 44.5\% of respondents, agreement with R3 was higher than agreement with R1; for 8.8\%, it was lower (Wilcoxon signed-rank test, p < 0.001). This descriptive difference does not by itself establish that the outcomes behave differently in the model. Section~\ref{sec:results} tests that question by comparing their relationships with the explanatory constructs.

\subsection{Explanatory constructs and indicators}\label{sec:constructs}

Twelve attitudinal statements measure three explanatory constructs and a separate control variable for satisfaction with current charging (Table~\ref{tab:items}). Respondents used the same five-point agreement scale as for the outcome questions.

\begin{itemize}
\item Perceived usefulness (PU1--PU4) captures four situations in which swapping might help: insufficient urban charging infrastructure, limited depot-based fast charging, slow public charging and concern about running out of range. In BRT terms, these statements represent reasons for swapping because they describe circumstances in which it may offer an advantage over plug-in charging.
\item Reasons against swapping (RA1--RA4) cover four practical barriers, following \citet{RamSheth1989}: too few stations, the cost of keeping charged batteries in reserve, differences in battery size and chemistry across manufacturers, and uncertainty about whether a compatible battery will be available. These items represent the barriers identified in Section~\ref{sec:lit-acceptance}: station coverage, operator cost, interoperability and battery availability.
\item Perceived energy-system strain (PES1--PES3) is measured with three statements about whether electricity demand is growing faster than generation and network capacity can accommodate. Because these statements concern the electricity system rather than battery swapping itself, we treat energy-system strain as a separate construct rather than a reason for or against swapping. We do not specify a directional hypothesis for it.
\item Status-quo satisfaction (SAT) captures whether respondents are content with their current charging options and therefore see less need for swapping. It is modelled separately from the reasons-against construct because it describes the existing alternative, not a barrier within swapping. This follows research that treats satisfaction with the status quo as a distinct source of resistance \citep{Heidenreich2015,Long2025}.
\end{itemize}

The descriptive results point to a central feature of the responses: people often recognised both the benefits and the barriers of swapping. The most widely endorsed statement concerned the difficulty of standardising batteries (RA3); the next most widely endorsed concerned the usefulness of swapping where urban charging infrastructure is insufficient (PU1). Respondents were therefore not simply positive or negative about the technology. Agreement was lower for the adoption outcomes than for the usefulness statements, and personal purchase preference received the least support.

\begin{table}
\caption{Outcome and indicator statements with descriptive statistics (N = 728)}\label{tab:items}
\small
\begin{tabularx}{\linewidth}{@{}lXrrr@{}}
\toprule
Code & Statement & Mean & SD & \% agree \\
\midrule
\multicolumn{5}{@{}l}{\textit{Own purchase preference (outcome)}} \\
R1 & I prefer to buy electric vehicles if battery swapping facility is available in general & 3.33 & 1.08 & 45.5 \\
\addlinespace
\multicolumn{5}{@{}l}{\textit{Expected system role (outcome)}} \\
R2 & Battery swapping potential would increase the electric vehicle uptake & 3.58 & 1.03 & 58.4 \\
R3 & Battery swapping could play a crucial role in shaping the future of transportation & 3.80 & 0.96 & 71.0 \\
\addlinespace
\multicolumn{5}{@{}l}{\textit{Perceived usefulness}} \\
PU1 & Battery swapping is useful when there is insufficient charging infrastructure in our cities & 3.98 & 0.96 & 78.7 \\
PU2 & Battery swapping is useful when there is a deficiency of depot based fast charging & 3.94 & 0.94 & 77.1 \\
PU3 & Battery swapping is useful as charging in public locations takes a long time & 3.78 & 1.17 & 68.3 \\
PU4 & Battery swapping is useful to reduce range anxiety & 3.81 & 1.11 & 68.3 \\
\addlinespace
\multicolumn{5}{@{}l}{\textit{Reasons against}} \\
RA1 & Insufficient availability of battery swapping stations limits the practicality of battery switching to reduce range anxiety and increase the driving range & 3.95 & 0.93 & 70.1 \\
RA2 & Holding a sufficient inventory of fully charged batteries for swapping results in additional expenses, increasing the overall costs of operations & 3.84 & 0.90 & 66.2 \\
RA3 & With different electric vehicle makers using different battery chemistries and sizes, creating a universal battery swap system remains difficult & 4.20 & 0.86 & 81.7 \\
RA4 & I am not sure I may find right batteries to swap at the battery swapping locations & 3.70 & 1.01 & 63.5 \\
\addlinespace
\multicolumn{5}{@{}l}{\textit{Perceived energy-system strain}} \\
PES1 & Increased electric mobility leads to a surge in electricity demand, which can be challenging for countries that intend to phase out nuclear energy and shift towards renewable energy sources & 3.44 & 1.00 & 49.6 \\
PES2 & Renewable energy sources alone may not satisfy the growing need for electricity, and those peak loading periods are unlikely to provide access to affordable electricity & 3.49 & 1.04 & 58.0 \\
PES3 & The current power network system may not be adequate to charge a commercial fleet; hence, upgrading the grid demands investment & 3.89 & 0.94 & 73.1 \\
\addlinespace
\multicolumn{5}{@{}l}{\textit{Status-quo satisfaction}} \\
SAT & I am happy with my current charging options and therefore battery swapping may not be useful & 3.57 & 1.24 & 58.9 \\
\bottomrule
\end{tabularx}

\vspace{3pt}
\begin{minipage}{\linewidth}\footnotesize \textit{Note:} Responses use a five-point scale from 1 (strongly disagree) to 5 (strongly agree). The percentage agreeing includes responses of 4 or 5.\end{minipage}
\end{table}

\subsection{Covariates}\label{sec:covariates}

The covariates enter the model in two ways (Table~\ref{tab:covariates}). Seven variables describing respondents and their charging circumstances help explain perceived usefulness, energy-system strain, reasons against swapping and expected system role. Five variables also enter the personal-preference equation directly, accounting for charging context and satisfaction with current options. Dependence on public charging appears in both parts because it may influence perceptions as well as purchase preference. The class-membership model (Section~\ref{sec:lca-method}) uses the same variables and adds satisfaction with local public-charging availability.

\begin{table}
\caption{Covariates: definitions, sample values and where each enters the model}\label{tab:covariates}
\small
\begin{tabularx}{\linewidth}{@{}P{0.2}XP{0.2}P{0.21}@{}}
\toprule
Covariate & Definition & Sample value & Enters \\
\midrule
Female & 1 if female, 0 otherwise & 0.514 & Latent constructs; class membership \\
Age & Seven ordered age bands, standardised & Most common: 35 to 44 & Latent constructs; class membership \\
Household income & Ordered household income bands, standardised & Most common: \pounds{}100,000 or more & Latent constructs; class membership \\
EV experience & Four ordered bands from less than 6 months to more than 3 years, standardised & Most common: 1 to 3 years & Latent constructs; class membership \\
Vehicle range & Six ordered electric-range bands from less than 100 km to more than 500 km, standardised & Most common: 200 to 300 km & Latent constructs; class membership \\
No home charging & 1 if the respondent does not charge at home & 0.152 & Latent constructs; class membership \\
Public-charging dependence & 1 if the respondent charges publicly weekly or daily & 0.283 & Latent constructs; own preference; class membership \\
Status-quo satisfaction & SAT, standardised & Mean 3.57 before standardisation & Own preference \\
Commuting & 1 if the EV is used for commuting & 0.787 & Own preference; class membership \\
Regional charge points, total\textsuperscript{*} & Public charge points per 100,000 population in the respondent's region, April 2025, standardised & 66.2 to 262.6 across regions & Own preference; class membership \\
Regional charge points, rapid\textsuperscript{*} & Rapid charge points per 100,000 population in the respondent's region, April 2025, standardised & 15.4 to 34.0 across regions & Own preference; class membership \\
Satisfaction with public charging & Five-point satisfaction with the availability of public charging in the area, standardised & Mean 3.15 before standardisation & Class membership \\
\bottomrule
\end{tabularx}

\vspace{3pt}
\begin{minipage}{\linewidth}\footnotesize \textsuperscript{*}Regional values are taken from the \citet{DfT2025Charging}.\end{minipage}
\end{table}

Satisfaction with current charging and the regional charger-supply measures are included as controls. Because the regional measures vary across only eleven regions, their coefficients should not be read as direct estimates of the effect of local infrastructure.

\section{Methodology}\label{sec:method}
This section sets out the modelling approach. Section~\ref{sec:iclv} specifies the integrated choice and latent variable model, Section~\ref{sec:identification} explains how the latent constructs are identified and scaled, Section~\ref{sec:standardisation} describes how effects on the two outcomes are standardised and compared, and Section~\ref{sec:lca-method} presents the latent class analysis used to identify groups of EV users.
\subsection{Integrated choice and latent variable model specification}\label{sec:iclv}

\begin{figure}
\centering
\includegraphics[width=\linewidth]{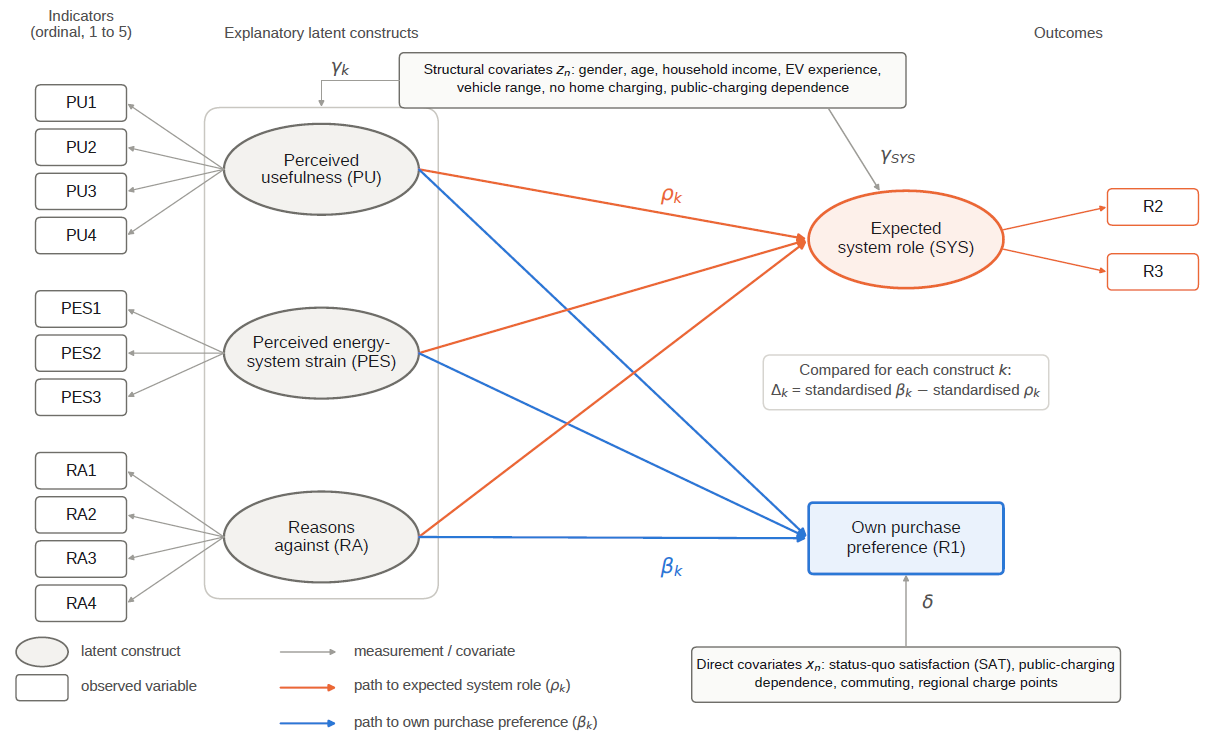}
\caption{Structure of the integrated choice and latent variable (ICLV) model.}\label{fig:iclv}
\end{figure}

We estimate an integrated choice and latent variable (ICLV) model \citep{BenAkiva2002,Bolduc2005,WalkerBenAkiva2002}. The model estimates four latent constructs and the outcome relationships together, in a single step. We do not create separate construct scores and then use them as if they were measured without error. This avoids one limitation of factor scores and summed indicators in the outcome model \citep{Daly2012,VijWalker2016}. Figure~\ref{fig:iclv} summarises the model.

Let n index respondents (n = 1, \ldots{}, N, with N = 728), and let j index the survey indicators (j = 1, \ldots{}, J, with J = 13). Each response has M = 5 ordered categories. The vector $z_n$ contains covariates used to explain the latent constructs, while $x_n$ contains covariates entered directly into the outcome equation. We model three explanatory constructs: perceived usefulness (PU), perceived energy-system strain (PES) and reasons against swapping (RA). Each is expressed as a linear function of the covariates plus a normally distributed disturbance:

\begin{equation}
LV_{kn} = \gamma_{k}' z_{n} + \sigma_{k}\eta_{kn}, \qquad \eta_{kn} \sim N(0,1), \qquad k \in \{PU, PES, RA\}
\label{eq:latent}
\end{equation}

In Equation~\eqref{eq:latent}, $LV_{kn}$ is the value of latent construct k for respondent n. The vector $z_n$ contains the respondent's demographic and charging-behaviour covariates (Section~\ref{sec:covariates}), and $\gamma_k$ contains the corresponding coefficients. The parameter $\sigma_k$ measures the scale of unexplained variation in that construct. The disturbance $\eta_{kn}$ is standard normal before it is multiplied by this scale, and is assumed to be independent across respondents and constructs.

Expected system role is modelled as an outcome of the three explanatory constructs and the covariates:

\begin{equation}
LV_{SYS,n} = \gamma_{SYS}' z_{n} + \sum_{k}\rho_{k} LV_{kn} + \sigma_{SYS}\,\eta_{SYS,n}
\label{eq:system}
\end{equation}

Here, $LV_{SYS,n}$ denotes respondent n's expected system role. The vector $\gamma_{SYS}$ contains the covariate coefficients, and $\rho_k$ gives the relationship between construct k and expected system role. The terms $\sigma_{SYS}$ and $\eta_{SYS,n}$ represent the outcome's disturbance scale and standard-normal disturbance. This disturbance is separate from the disturbances associated with the three explanatory constructs.

Each latent construct is measured through its survey indicators using an ordered logit model. For an indicator assigned to construct k, the model estimates the probability of each response category as follows:

\begin{equation}
\Pr\left(I_{jn}=m \mid LV_{k(j)n}\right) = \Lambda\left(\tau_{jm}-\zeta_{j} LV_{k(j)n}\right) - \Lambda\left(\tau_{j,m-1}-\zeta_{j} LV_{k(j)n}\right)
\label{eq:measurement}
\end{equation}

Here, $y_{jn}$ is respondent n's answer to indicator j, $k_j$ identifies the construct measured by that indicator, $\zeta_j$ is the item loading, and $\tau_{jm}$ is the threshold for response category m. The logistic cumulative distribution function, $\Lambda$, converts the underlying construct value into probabilities for the observed response. The lowest and highest response categories have open-ended thresholds:

\begin{equation}
\Lambda(a) = \left[1+\exp(-a)\right]^{-1}, \qquad \tau_{j0}=-\infty, \qquad \tau_{jM}=+\infty
\label{eq:logistic}
\end{equation}

The logistic cumulative distribution function is defined for any real value. Each indicator has its own estimated thresholds, with the lowest and highest categories left open-ended. To keep the thresholds in order, we estimate the first threshold freely and express each later threshold as the previous threshold plus a positive increment:

\begin{equation}
\tau_{j1}=\alpha_{j1}, \qquad \tau_{jm}=\tau_{j,m-1}+\exp\left(\alpha_{jm}\right), \qquad m=2,\ldots,M-1
\label{eq:thresholds}
\end{equation}

The first threshold, $\alpha_{j1}$, is estimated freely for each indicator. Each later threshold is formed by adding an exponentiated increment. Because an exponential is always positive, the thresholds remain correctly ordered throughout estimation without separate inequality constraints.

Personal purchase preference is modelled using a second ordered logit equation. Its underlying index depends on the latent constructs and on covariates entered directly into this outcome equation:

\begin{equation}
V_{n} = \sum_{k}\beta_{k} LV_{kn} + \delta' x_{n}
\label{eq:index}
\end{equation}

Here, $V_n$ is the underlying index for respondent n's personal purchase preference, and $\beta_k$ is the coefficient for latent construct k. The vector $x_n$ contains the covariates entered directly into this equation, with $\delta$ denoting their coefficients. These include satisfaction with current charging, dependence on public charging, commuting and the two regional charger-supply measures. The probability of each observed response is then calculated as follows:

\begin{equation}
\Pr\left(y_{n}=m \mid LV_{n}\right) = \Lambda\left(c_{m}-V_{n}\right) - \Lambda\left(c_{m-1}-V_{n}\right)
\label{eq:preference}
\end{equation}

The observed response $y_n$ is the respondent's answer to R1 on the five-point scale. The cut-points $c_m$ are estimated using the positive-increment approach in Equation~\eqref{eq:thresholds}. As with the indicator equations, the lowest and highest categories are open-ended, so $c_0$ and $c_M$ are unbounded.

Expected system role does not predict personal purchase preference in this specification. The two are parallel outcomes, not a predictor and mediator. Because both depend on the same explanatory constructs, we can compare how strongly each construct relates to each outcome without assuming that one outcome causes the other.

Conditional on the latent constructs, the indicator responses and personal-preference response are assumed to be independent. The likelihood for each respondent is found by integrating their joint probabilities over the four disturbance terms:

\begin{equation}
L_{n}(\theta) = \int_{\mathbb{R}^{4}} \left[\prod_{j=1}^{13}\Pr\left(I_{jn}\mid\eta\right)\right]\Pr\left(y_{n}\mid\eta\right)\,\phi(\eta)\,d\eta
\label{eq:likelihood}
\end{equation}

The parameter vector $\theta$ contains all estimated parameters, $\eta$ is the vector of four disturbances, and $\phi$ is the standard multivariate normal density with identity covariance. The responses are correlated overall because they share latent constructs, even though they are assumed independent once those constructs are given. This shared structure is why the model is estimated jointly. The integral has no closed-form solution, so we approximate it by simulation:

\begin{equation}
\widehat{L}_{n}(\theta) = \frac{1}{R}\sum_{r=1}^{R}\left[\prod_{j=1}^{13}\Pr\left(I_{jn}\mid\eta^{(r)}\right)\right]\Pr\left(y_{n}\mid\eta^{(r)}\right)
\label{eq:simlik}
\end{equation}

Here, R is the number of simulation draws and $\eta^{(r)}$ is the r-th draw from the standard multivariate normal distribution. We keep the draws fixed across estimation iterations and maximise the sum of respondents' log-likelihood contributions.

Estimation is carried out in Apollo version 0.3.9 using MLHS draws, the BGW optimiser, analytical gradients and robust standard errors \citep{HessPalma2019}. The four disturbance terms use independent draw sequences; their realised absolute correlations are below 0.01. This independence is important. If the same sequence were used for every dimension, the four-dimensional integral would effectively collapse to one dimension, leaving the latent constructs unidentified separately.

\subsection{Identification and normalisation}\label{sec:identification}

A latent construct needs a defined scale. Two common approaches are to fix one indicator loading at one and estimate the disturbance scale, or to fix the disturbance scale at one and estimate all loadings \citep{Daly2012}. We use the first approach. When the disturbance scale is estimated freely, the likelihood is unchanged by the joint rescaling shown in Equation~\eqref{eq:rescaling}:

\begin{equation}
\left(LV_{k},\ \zeta_{j},\ \gamma_{k},\ \sigma_{k},\ \beta_{k},\ \rho_{k}\right) \mapsto \left(c\,LV_{k},\ \zeta_{j}/c,\ c\,\gamma_{k},\ c\,\sigma_{k},\ \beta_{k}/c,\ \rho_{k}/c\right), \qquad c>0
\label{eq:rescaling}
\end{equation}

The scalar c in this transformation can be any positive value. It rescales the construct and adjusts every parameter that multiplies it or is multiplied by it, while leaving the response probabilities in Equations~\eqref{eq:measurement} and~\eqref{eq:preference} unchanged. Fixing one indicator loading, $\zeta_j$, at one therefore defines the scale; it does not impose an additional restriction on the data.

Fixing both an indicator loading and the disturbance scale is not simply another way to set the scale; it adds a restriction that should be tested. Equation~\eqref{eq:rescaling} requires the disturbance scale to change when the construct is rescaled. If that scale is fixed at one, this adjustment is no longer possible. The loadings are then fixed in absolute terms, and setting a loading to one becomes a testable assumption rather than a normalisation. The logic is similar to item-response models, where equal item discrimination is a testable restriction rather than a scaling convention.

The data reject this additional restriction. Imposing it reduces the log-likelihood by 150.22 points across the four constructs, giving a likelihood-ratio statistic of 300.44 on four degrees of freedom (p < 0.001). The disturbance scales for perceived usefulness and energy-system strain are far from one, whereas those for reasons against and expected system role are closer to one. We therefore do not impose both scaling conventions at once: doing so would add an assumption that is inconsistent with these data.

\subsection{Standardisation and comparison of effects across outcomes}\label{sec:standardisation}

Personal purchase preference and expected system role are measured on different scales. The ordered logit for personal preference has an underlying error variance of $\pi^2/3$, whereas expected system role has an estimated disturbance scale. Their coefficients, $\beta_k$ and $\rho_k$, cannot therefore be compared directly. We standardise the effects using the dispersion of each outcome, with the relevant variances defined below:

\begin{equation}
\operatorname{Var}\left(LV_{k}\right) = \operatorname{Var}_{n}\left(\gamma_{k}' z_{n}\right) + \sigma_{k}^{2}
\label{eq:varlv}
\end{equation}

\begin{equation}
\operatorname{Var}\left(LV_{SYS}\right) = \operatorname{Var}_{n}\left(\gamma_{SYS}' z_{n} + \sum_{k}\rho_{k}\gamma_{k}' z_{n}\right) + \sum_{k}\rho_{k}^{2}\sigma_{k}^{2} + \sigma_{SYS}^{2}
\label{eq:varsys}
\end{equation}

\begin{equation}
\operatorname{Var}\left(y_{n}^{*}\right) = \operatorname{Var}_{n}\left(\sum_{k}\beta_{k}\gamma_{k}' z_{n} + \delta' x_{n}\right) + \sum_{k}\beta_{k}^{2}\sigma_{k}^{2} + \frac{\pi^{2}}{3}
\label{eq:vary}
\end{equation}

Here, $\operatorname{Var}_n$ is the variance across respondents of the predicted quantity, evaluated at the estimated parameters. The latent propensity $y_n^{*}$ is the unobserved value underlying the ordered response, while $\pi^2/3$ is the variance of the logistic error in Equation~\eqref{eq:preference}. Each variance includes variation explained by covariates and variation contributed by the disturbance terms.

For each construct, we compare its standardised relationship with personal purchase preference against its standardised relationship with expected system role:

\begin{equation}
\Delta_{k} = \frac{\beta_{k}\operatorname{SD}\left(LV_{k}\right)}{\operatorname{SD}\left(y^{*}\right)} - \frac{\rho_{k}\operatorname{SD}\left(LV_{k}\right)}{\operatorname{SD}\left(LV_{SYS}\right)}
\label{eq:contrast}
\end{equation}

The contrast $\Delta_k$ is the difference between a construct's standardised relationship with personal purchase preference and its standardised relationship with expected system role. A negative contrast indicates a stronger relationship with expected system role; a positive contrast indicates a stronger relationship with personal preference. We calculate standard errors using the delta method and the robust covariance matrix:

\begin{equation}
\widehat{\operatorname{Var}}\left(\Delta_{k}\right) = \left(\frac{\partial\Delta_{k}}{\partial\theta}\right)'\,\widehat{\Sigma}\,\left(\frac{\partial\Delta_{k}}{\partial\theta}\right)
\label{eq:deltamethod}
\end{equation}

$\Sigma_{\theta}$ is the robust covariance matrix of the estimated parameters, and the gradient of $\Delta_k$ is calculated numerically around the estimates. The contrast depends on 43 model parameters; most of the remaining parameters, including the response thresholds, do not enter its calculation. Both the contrast and its standard error are unchanged by the rescaling in Equation~\eqref{eq:rescaling}, although individual loading test statistics can change. We therefore use the contrasts to compare the two outcomes and use loading significance only to assess whether each indicator contributes to its intended construct.

Hypothesis H3 tests whether the two asymmetries differ. Reasons against swapping have negative relationships with both outcomes, so a stronger relationship with personal preference corresponds to a negative contrast for that construct. We reverse the sign of this contrast and add it to the usefulness contrast:

\begin{equation}
D = -\left(\Delta_{PU} + \Delta_{RAGN}\right)
\label{eq:dstat}
\end{equation}

The test statistic D is positive when usefulness is more strongly related to expected system role and reasons against are more strongly related to personal preference. A value of zero means that the two constructs divide their influence between the outcomes in the same proportion. We calculate its standard error using the same delta method. We also report the ratio $R_k$ of each construct's two standardised effects:

\begin{equation}
R_{k} = \frac{\beta_{k}\operatorname{SD}\left(LV_{k}\right)/\operatorname{SD}\left(y^{*}\right)}{\rho_{k}\operatorname{SD}\left(LV_{k}\right)/\operatorname{SD}\left(LV_{SYS}\right)}
\label{eq:ratio}
\end{equation}

A ratio $R_k$ below one indicates a stronger relationship with expected system role, while a ratio above one indicates a stronger relationship with personal purchase preference. Comparing ratios across constructs removes any common scaling factor affecting personal preference. The comparison is therefore robust to measurement error in the single-item preference measure, as explained in Section~\ref{sec:structural}.

\subsection{Latent class segmentation and distal outcome analysis}\label{sec:lca-method}

We estimate the latent class model separately from the outcome model, using the twelve attitudinal items and excluding the three adoption-potential items. The outcomes are therefore not used to define the classes. For a model with C classes, the likelihood of respondent n's full set of answers is:

\begin{equation}
\Pr\left(I_{n}\right) = \sum_{c=1}^{C}\pi_{c}\prod_{j=1}^{12}\prod_{m=1}^{5}\vartheta_{cjm}^{\,\mathbb{1}\left[I_{jn}=m\right]}
\label{eq:lca}
\end{equation}

Here, $\pi_c$ is the share of respondents in class c, and the class shares sum to one. The response probability $\vartheta_{cjm}$ is the chance that a member of class c selects category m for item j. The indicator function takes a value of one when respondent n gives that response and zero otherwise.

We estimate solutions with one to six classes using the EM algorithm, with ten random starting points for each solution. BIC guides the choice of class number; we also examine entropy and the size of the smallest class to judge how clearly the groups are separated and whether each group is useful to interpret. We describe the selected classes in three ways. First, we summarise the twelve survey indicators using class-specific means implied by the estimated response probabilities, which equal posterior-weighted sample means at convergence. Second, we report posterior-weighted means of the covariates for each class.

Because the three adoption outcomes were not used to create the classes, we analyse them separately using the BCH correction \citep{Bolck2004,Vermunt2010}. Assigning each respondent to their most likely class can misclassify people and make class averages look more alike than they are. BCH uses each respondent's estimated classification error to adjust the outcome comparisons. We calculate, for each class, the gap between system-level judgement and personal judgement:

\begin{equation}
G_{c} = \mu_{c}\left(\mathrm{R3}\right) - \mu_{c}\left(\mathrm{R1}\right)
\label{eq:gap}
\end{equation}

Here, $\mu_c$ is the BCH-corrected mean for class c. The difference between two classes' gaps, $G_a - G_b$, is a difference-in-differences: it accounts for general differences in agreement between the groups that affect both outcomes. Confidence intervals and two-sided p-values come from 3,000 bootstrap samples of respondents, with the weights recalculated in every sample.

We use a multinomial logit model with the maximum-likelihood three-step correction to examine how covariates relate to class membership \citep{Vermunt2010}. This method accounts for classification error while keeping the covariates out of the measurement model. The largest class serves as the reference category. We estimate classes from the original ordered responses rather than composite scores, because averages of three or four five-point items take only a limited number of values. A continuous mixture model can then group repeated score values rather than identify meaningful differences between respondents (Section~\ref{sec:lca-selection}).

\section{Results}\label{sec:results}

Section~\ref{sec:fit} reports overall model fit and checks whether the survey indicators measure their intended constructs. Sections~\ref{sec:structural} and~\ref{sec:items} examine the relationships with the two outcomes, first at construct level and then for each barrier item. Section~\ref{sec:lca} describes the user groups, and Section~\ref{sec:hyp-summary} brings the hypothesis tests together.

\subsection{Model fit and measurement model}\label{sec:fit}

The joint model estimates 108 free parameters and reaches a log-likelihood of $-$11,992.29 across 10,192 ordinal responses, with fourteen responses modelled per respondent. The Hessian is negative definite at the solution. Against a null model that assigns equal probability to every response category, the model has a $\rho^2$ of 0.269 and an adjusted $\rho^2$ of 0.262. Against a thresholds-only model that reproduces the observed distribution of each item, $\rho^2$ is 0.113. The likelihood-ratio statistic is 3,054.7 on 52 degrees of freedom (p < 0.001). The AIC is 24,200.58 and the BIC is 24,696.33.

All thirteen indicators are significantly related to their assigned constructs (Table~\ref{tab:loadings}). Even the weakest relationship is statistically significant (p = 0.005).

\begin{table}
\caption{Measurement-model loadings and robust standard errors}\label{tab:loadings}
\small
\begin{tabularx}{\linewidth}{@{}Xlrrrr@{}}
\toprule
Construct & Indicator & Loading & Robust SE & t & p \\
\midrule
Perceived usefulness & PU1 & 1.000 & fixed for scale &  &  \\
 & PU2 & 1.121 & 0.134 & 8.36 & < 0.001 \\
 & PU3 & 0.532 & 0.080 & 6.64 & < 0.001 \\
 & PU4 & 0.485 & 0.074 & 6.57 & < 0.001 \\
Energy-system strain & PES1 & 1.000 & fixed for scale &  &  \\
 & PES2 & 1.062 & 0.169 & 6.28 & < 0.001 \\
 & PES3 & 0.802 & 0.197 & 4.08 & < 0.001 \\
Reasons against & RA2 & 1.000 & fixed for scale &  &  \\
 & RA3 & 1.370 & 0.350 & 3.91 & < 0.001 \\
 & RA4 & 1.220 & 0.434 & 2.81 & 0.005 \\
 & RA1 & 0.750 & 0.210 & 3.57 & < 0.001 \\
Expected system role & R2 & 1.000 & fixed for scale &  &  \\
 & R3 & 1.236 & 0.160 & 7.72 & < 0.001 \\
\bottomrule
\end{tabularx}
\end{table}

The estimated disturbance scales are 4.05 for perceived usefulness and 1.65 for energy-system strain; both differ significantly from one. The corresponding scales for reasons against and expected system role are 0.37 and 0.63, neither significantly different from one. The larger scales for usefulness and energy-system strain help explain why the joint restriction described in Section~\ref{sec:identification} was rejected.

Reasons against is the least precisely measured construct, so we checked whether adding another indicator improved the measurement. With three indicators, its disturbance scale was 0.25 and the standard error of its standardised contrast was 0.298. Including RA1 as a fourth indicator raised the scale to 0.37 and reduced the standard error to 0.146, while leaving the estimated effects broadly similar. We report both specifications below.

\subsection{Structural model: standardised effects on the two outcomes}\label{sec:structural}

\begin{figure}
\centering
\includegraphics[width=\linewidth]{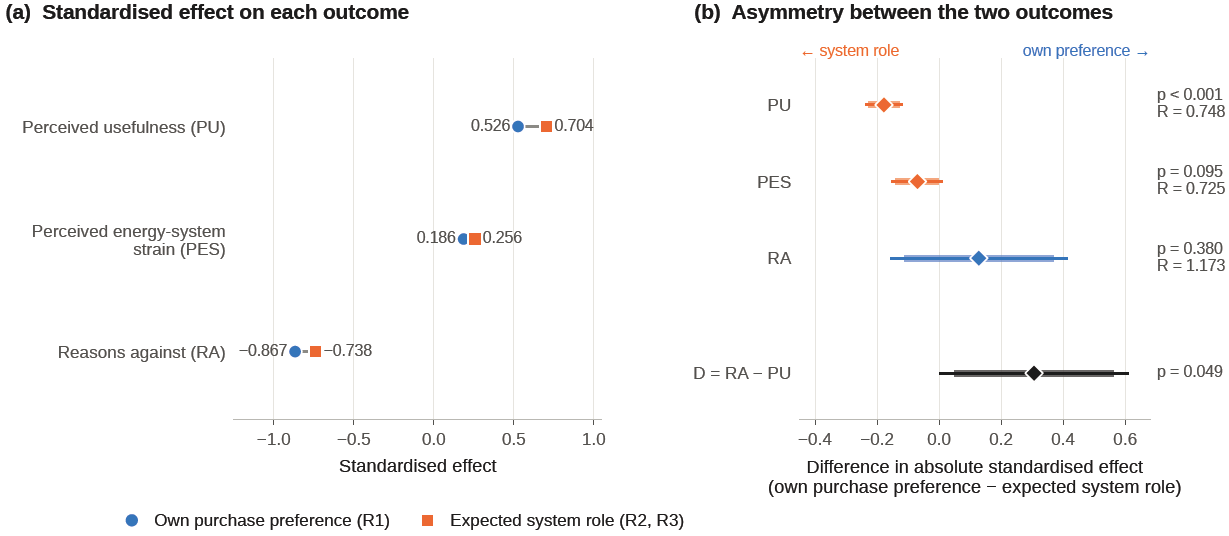}
\caption{Standardised relationships between the three explanatory constructs and personal purchase preference and expected system role. Panel (a) shows each outcome separately. Panel (b) compares the absolute standardised effects and reports 90\% and 95\% confidence intervals calculated using the delta method. R is the ratio of the two effects, and D is the test statistic for H3.}\label{fig:stdeffects}
\end{figure}

Perceived usefulness and perceived energy-system strain have stronger relationships with expected system role than with personal purchase preference. Reasons against swapping point in the opposite direction, with a stronger relationship to personal preference (Figure~\ref{fig:stdeffects}).

A ratio below one indicates that a construct is more strongly related to expected system role; a ratio above one indicates a stronger relationship with personal preference. The difference between the usefulness and reasons-against patterns is 0.306 (SE 0.155, z = 1.97, p = 0.049). Excluding RA1 gives estimates of $-$0.190, $-$0.070 and $-$0.202, with a difference of 0.392 (SE 0.307, p = 0.202). The direction of the findings remains similar, but the difference is estimated less precisely without RA1.

On its own, the reasons-against contrast is not statistically significant in either specification. We therefore do not interpret its estimated size in isolation. Instead, we consider its direction alongside the joint test and the item-level results in Section~\ref{sec:items}, which reveal clearer differences among the barriers.

Personal purchase preference is measured with one question, whereas expected system role is a latent outcome measured by two questions and adjusted for measurement error. Random measurement error in R1 would reduce all standardised effects on personal preference by a common factor. It would therefore make the apparent reasons-against asymmetry smaller, not larger, while leaving the comparison between constructs unchanged because the ratio of their R values (1.173 / 0.748 = 1.568) is unaffected. Correcting for this error would increase D. For usefulness, a conservative correction based on the standardised loading of R1 (0.821) in a confirmatory factor model of the three adoption questions reduces the contrast to $-$0.063, but its direction remains unchanged. Table~\ref{tab:paths} reports the structural paths on their original scales, together with robust standard errors.

\begin{table}
\caption{Structural paths on their native scales}\label{tab:paths}
\small
\begin{tabularx}{\linewidth}{@{}Xrrrr@{}}
\toprule
Path & Estimate & Robust SE & t & p \\
\midrule
Usefulness to expected system role & 0.418 & 0.071 & 5.86 & < 0.001 \\
Usefulness to own preference & 0.561 & 0.193 & 2.91 & 0.004 \\
Energy strain to expected system role & 0.376 & 0.174 & 2.17 & 0.030 \\
Energy strain to own preference & 0.489 & 0.179 & 2.74 & 0.006 \\
Reasons against to expected system role & $-$3.954 & 2.932 & $-$1.35 & 0.178 \\
Reasons against to own preference & $-$8.327 & 9.674 & $-$0.86 & 0.389 \\
\bottomrule
\end{tabularx}
\end{table}

The reasons-against coefficients are large in part because they are divided by a small disturbance scale of 0.37. Their magnitude therefore reflects the scale of the latent construct and should not be read directly as evidence of a strong relationship. Neither path is statistically significant on its own. By contrast, the paths from usefulness and energy-system strain are significant for both outcomes, so the comparison in Figure~\ref{fig:stdeffects} is not driven by an imprecisely estimated relationship on one side.

Five relationships between the covariates and latent constructs are statistically significant at the 10\% level.

\begin{table}
\caption{Covariate effects on the latent constructs significant at the 10\% level}\label{tab:covlatent}
\small
\begin{tabularx}{\linewidth}{@{}Xrrr@{}}
\toprule
Effect & Estimate & Robust SE & p \\
\midrule
Public-charging dependence on reasons against & $-$0.571 & 0.133 & < 0.001 \\
Vehicle range on perceived usefulness & $-$0.560 & 0.194 & 0.004 \\
Female on perceived usefulness & 0.885 & 0.400 & 0.027 \\
Vehicle range on energy-system strain & $-$0.201 & 0.101 & 0.047 \\
Age on energy-system strain & 0.167 & 0.089 & 0.061 \\
\bottomrule
\end{tabularx}
\end{table}

Two covariate findings stand out. Drivers of longer-range EVs see less usefulness in swapping, as expected if they have less need for a solution to limited range. Longer range is also associated with lower perceived pressure on the electricity system. Respondents who rely on public charging report fewer reasons against swapping, consistent with swapping being an alternative to public charging rather than home charging. The other 23 covariate terms are not significant. In this model, satisfaction with current charging, commuting and the two regional charger-supply measures also have no significant direct relationship with personal preference.

\subsection{Item-level analysis of reasons against}\label{sec:items}

\begin{figure}
\centering
\includegraphics[width=\linewidth]{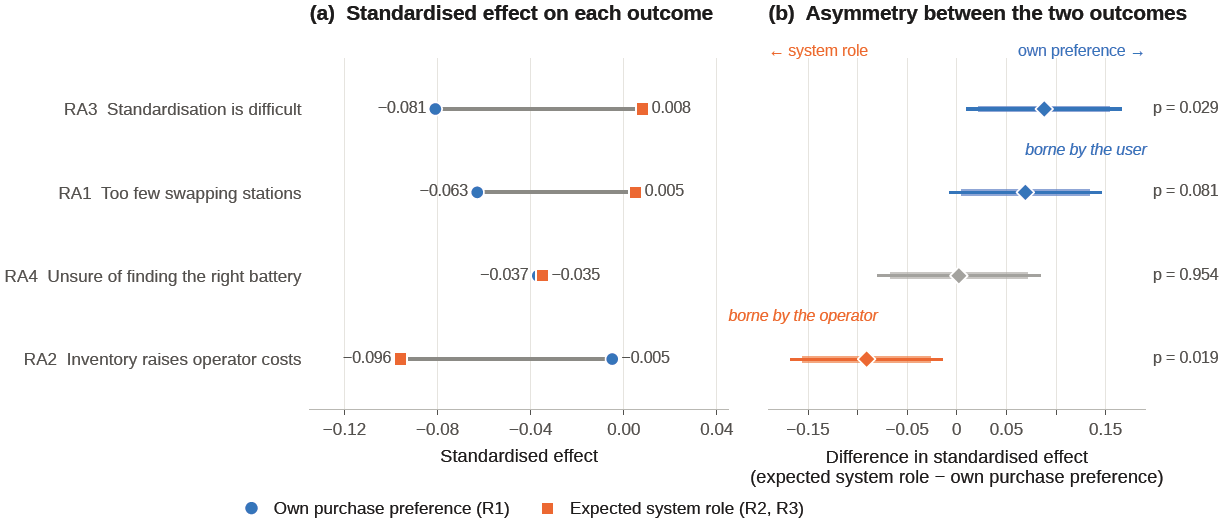}
\caption{Relationships between the four reasons-against items and the two outcomes when each item is entered separately as an observed ordinal variable. Panel (a) shows the standardised effect of each item on each outcome. Panel (b) shows the difference between the two effects, with 90\% and 95\% confidence intervals. Positive values indicate a more negative relationship with personal purchase preference. The joint Wald test that all four differences equal zero is $\chi^2(4)$ = 12.82, p = 0.012.}\label{fig:itemlevel}
\end{figure}

Entering the four barrier items separately as observed ordinal variables, rather than combining them into one latent construct, reveals a clearer crossover between the outcomes (Figure~\ref{fig:itemlevel}). The items differ in both the direction and size of their relationships, and the estimates are more precise than those for the overall construct. The largest standard error in this model is 1.38, compared with 11.36 when the barriers are treated as latent. Because the observed items have their own scales, their coefficients are not confounded with an estimated scale for a latent construct.

The joint Wald test rejects the hypothesis that all four item-level differences are zero ($\chi^2$ = 12.82, 4 degrees of freedom, p = 0.012). The pattern depends on who bears the cost or inconvenience. Operators must pay to keep spare batteries in reserve, so inventory cost appears to shape expectations about whether a swapping network can operate. Drivers, in contrast, are directly affected by incompatible batteries and limited station coverage because these determine whether they can use the network. Both concerns are more closely related to personal purchase preference. The remaining item, uncertainty about whether the right battery will be available, is less conclusive and is not significantly related to either outcome.

This interpretation was not built into the model from the outset; it emerged through specification checks. RA1 was first excluded from the reasons-against construct on conventional measurement grounds, then restored when the measurement model was estimated jointly using ordered logit equations. We examined it both as a fourth indicator of the latent construct and as an observed item in its own right. We had expected its focus on station availability, rather than on the driver, to place it alongside operator cost as a system-level concern. Instead, it behaved more like battery standardisation and showed almost no relationship with expected system role. This suggests that the key distinction may be who bears the consequences of a barrier, rather than how the item is worded. Future surveys should test this directly (Section~\ref{sec:conclusions}).

\subsection{Latent class segmentation of EV users}\label{sec:lca}

The joint model describes how reasons relate to the outcomes on average. The class analysis asks a different question: how do people combine these reasons? It identifies groups that differ in the benefits they recognise, the barriers they report and their satisfaction with current charging. It then examines whether these groups also differ in personal purchase preference and expected system role.

\subsubsection{Model selection}\label{sec:lca-selection}

The four-class solution has the lowest BIC (Table~\ref{tab:lcafit}). BIC falls substantially as the number of classes increases from one to four, then rises. Entropy is 0.851, above the commonly used 0.80 benchmark for clear separation. The smallest class contains 7.7\% of respondents (56 people). Classification accuracy ranges from 0.85 for the disengaged class to 0.97 for rejecters. Using each respondent's most likely class, the class shares are 26.5\%, 42.7\%, 23.1\% and 7.7\%; the posterior-weighted shares are 26.2\%, 41.5\%, 24.6\% and 7.7\%.

\begin{table}
\caption{Latent class solutions on the twelve attitudinal items}\label{tab:lcafit}
\small
\begin{tabularx}{\linewidth}{@{}Xrrrrr@{}}
\toprule
Classes & Parameters & Log-likelihood & BIC & Entropy & Smallest class \\
\midrule
1 & 48 & $-$11,585.1 & 23,486.5 & -- & 100.0\% \\
2 & 97 & $-$11,014.4 & 22,668.0 & 0.869 & 31.8\% \\
3 & 146 & $-$10,692.0 & 22,346.2 & 0.901 & 14.9\% \\
\textbf{4} & \textbf{195} & \textbf{$-$10,497.1} & \textbf{22,279.4} & \textbf{0.851} & \textbf{7.7\%} \\
5 & 244 & $-$10,359.1 & 22,326.2 & 0.862 & 6.5\% \\
6 & 293 & $-$10,251.7 & 22,434.4 & 0.866 & 5.9\% \\
\bottomrule
\end{tabularx}
\end{table}

We also tested a latent profile model based on composite scores, but did not use it for the main analysis. With a minimum-variance constraint, BIC selected two groups, but entropy was only 0.484, indicating weak separation. Without that constraint, a five-group model appeared preferable only because one class collapsed around a single usefulness score, with a standard deviation of 0.001. Each composite is the average of three or four five-point items and therefore has only eleven to seventeen possible values, some of which occur frequently. For example, usefulness was exactly 4.0 for 132 respondents and 5.0 for 104. The mixture model was fitting these repeated scores rather than clear population differences. The ordinal item-level model avoids this problem and respects the ordered nature of the survey responses.

\subsubsection{Class profiles on the attitudinal indicators}\label{sec:lca-profiles}

The four classes differ less in whether they recognise barriers than in whether they also see benefits and how satisfied they are with current charging (Figure~\ref{fig:classprofiles}; Table~\ref{tab:classprofiles} in Appendix~\ref{app:class-profiles}). Perceived usefulness provides the clearest separation: one group almost universally agrees that swapping would help, while another mostly disagrees. Battery standardisation, however, is widely seen as a barrier in every class.

\begin{figure}
\centering
\includegraphics[width=0.78\linewidth]{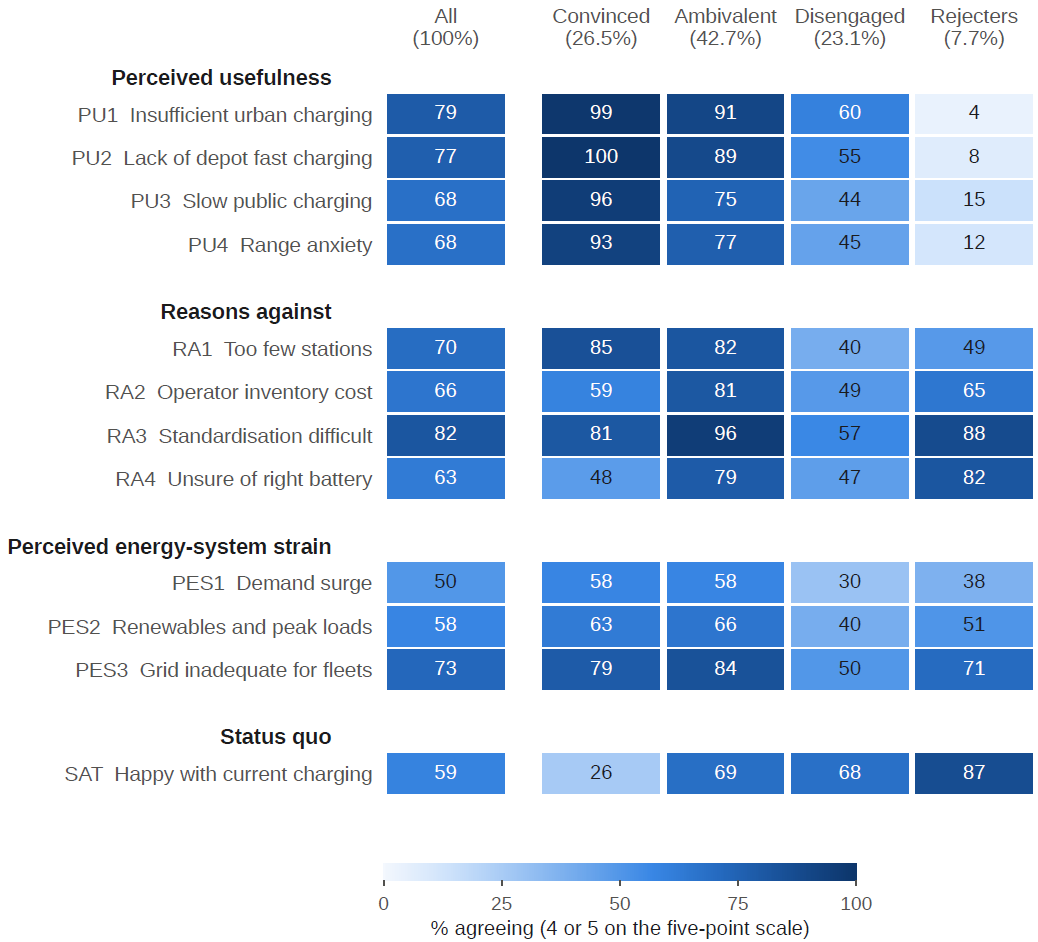}
\caption{Profiles of the four latent classes across the twelve attitudinal items, shown as the model-estimated share agreeing (responses 4 or 5).}\label{fig:classprofiles}
\end{figure}

The class labels summarise these response patterns and the differences in adoption outcomes presented in Section~\ref{sec:lca-outcomes}.

\begin{itemize}
\item Convinced (26.5\%). This group sees clear benefits in swapping: between 93\% and 100\% agree with the usefulness statements. Its members are also the least satisfied with current charging (26\%). Fewer express concern about operator costs (59\%) or finding the right battery (48\%) than in the other groups. Even so, they recognise network barriers: 85\% report too few stations and 81\% say battery standardisation is difficult.
\item Ambivalent (42.7\%). Most respondents in this group see the benefits of swapping (75\% to 91\% agree with the usefulness statements), but they also report the strongest concerns about barriers. Almost all identify battery standardisation as a problem (96\%); many worry about inventory costs (81\%) and whether the right battery will be available (79\%). Their perceived pressure on the electricity system is as high as in the convinced group. They see promise in swapping, but are also keenly aware of the practical obstacles.
\item Disengaged (23.1\%). Responses in this group tend to sit near the middle of the scale. No usefulness or reasons-against item receives more than 60\% agreement, and respondents select the neutral category for 38.9\% of their answers, nearly three times as often as people in the other groups. They also report the lowest perceived pressure on the electricity system. This pattern is more consistent with uncertainty or a lack of a settled view than with outright opposition.
\item Rejecters (7.7\%). Most members of this group do not see swapping as useful: only 4\% to 15\% agree with the usefulness statements. They are also the most satisfied with current charging (87\%). They strongly endorse barriers that affect users, including standardisation (88\%) and uncertainty about finding the right battery (82\%). Fewer than half, however, say that too few stations limit swapping. One possible explanation is that this item links station availability to reducing range anxiety, a benefit that this group does not recognise.
\end{itemize}

\subsubsection{Adoption outcomes by class}\label{sec:lca-outcomes}

\begin{figure}
\centering
\includegraphics[width=\linewidth]{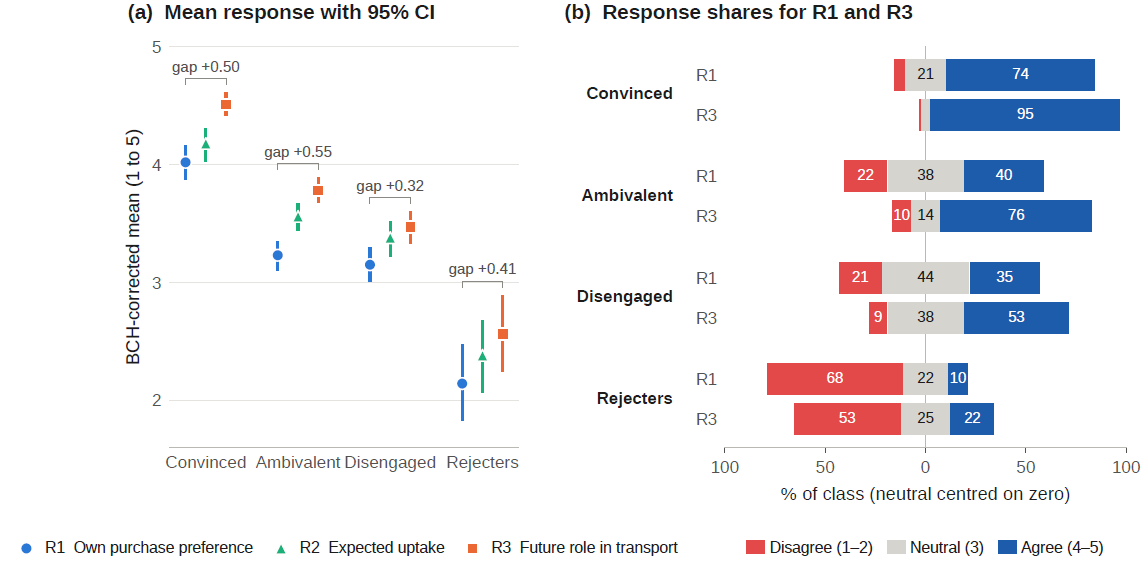}
\caption{Adoption outcomes by latent class. Panel (a) shows BCH-corrected mean responses to R1, R2 and R3, with 95\% bootstrap confidence intervals based on 3,000 resamples. Panel (b) shows BCH-weighted shares who disagree (responses 1 or 2), are neutral (response 3) or agree (responses 4 or 5) with R1 and R3; the neutral share is centred on zero.}\label{fig:classoutcomes}
\end{figure}

The three adoption outcomes rank the classes in the same order: convinced, ambivalent, disengaged and rejecters. Across all four groups, agreement is higher for statements about the system than for the statement about personal purchase preference (Figure~\ref{fig:classoutcomes}a).

Sixteen of the eighteen pairwise class differences across the three outcomes are statistically significant at p < 0.001. The two exceptions concern the ambivalent and disengaged groups. Their personal purchase preferences do not differ significantly (+0.08, p = 0.481), and the difference in expected uptake is also not significant at conventional levels (+0.18, p = 0.088). The classes are most clearly separated at the extremes: about three quarters of the convinced group would prefer a vehicle that supports swapping, compared with roughly one in ten rejecters.

The gap between system-level and personal judgements also varies across classes, broadly matching H1 and H2. The gap is largest in the ambivalent group (+0.55), whose members see substantial benefits but report the strongest barriers. It is smallest in the disengaged group (+0.32), whose members report less of both. One interpretation is that perceived usefulness raises expectations about the system, while barriers make personal adoption less attractive. Because this ordering compares only four classes, it is descriptive; the contrast between the ambivalent and disengaged groups provides a more direct test (Table~\ref{tab:ambdis}).

\begin{table}
\caption{Ambivalent minus disengaged class: differences in adoption outcomes}\label{tab:ambdis}
\small
\begin{tabularx}{\linewidth}{@{}Xrcr@{}}
\toprule
Quantity & Difference & 95\% CI & p \\
\midrule
Own purchase preference (R1) & +0.08 & $-$0.13 to +0.29 & 0.481 \\
Expected uptake (R2) & +0.18 & $-$0.02 to +0.38 & 0.088 \\
Future role (R3) & +0.31 & +0.13 to +0.50 & < 0.001 \\
Difference in gaps, future role minus own preference & +0.23 & +0.03 to +0.43 & 0.016 \\
Difference in gaps, expected uptake minus own preference & +0.10 & $-$0.10 to +0.30 & 0.334 \\
Difference in gaps, share agreeing (percentage points) & +18.2 & +5.4 to +30.7 & 0.007 \\
Difference in gaps, share disagreeing (percentage points) & $-$0.1 & $-$9.5 to +9.3 & 0.962 \\
\bottomrule
\end{tabularx}
\end{table}

The ambivalent and disengaged groups report similar personal purchase preferences, but they differ in what they expect from the wider system. After allowing for general differences in agreement between the groups, the difference between their outcome gaps is 0.23 (p = 0.016). The response shares help explain why. Similar proportions in both groups disagree that swapping could play a crucial future role (9.7\% and 9.4\%) or disagree with the personal-purchase statement (21.8\% and 21.4\%). The main difference is in agreement with the future-role statement: 76.0\% of the ambivalent group agree, compared with 52.6\% of the disengaged group; 38.0\% of the disengaged group remain neutral. The disengaged group is therefore not simply rejecting a future role for swapping. Many members remain undecided, while the ambivalent group is more likely to see a future role despite its concerns.

This comparison covers only one pair of classes. It supports the construct-level findings, but it is not an independent test of H1 to H3.

\subsubsection{Covariate profiles and class membership}\label{sec:lca-covariates}

Respondents' characteristics and charging habits help describe class membership, but they do not determine it (Table~\ref{tab:covprofiles}). The eleven covariates are jointly significant in the membership model (likelihood-ratio $\chi^2$ = 69.49 on 33 degrees of freedom, p = 0.0002). Yet no single covariate changes the odds of belonging to a class by more than about a factor of two (Table~\ref{tab:mnl}). Attitudes, rather than demographic or charging characteristics alone, therefore provide the main basis for distinguishing the groups.

\begin{table}
\caption{Covariate profiles by class (posterior-weighted)}\label{tab:covprofiles}
\small
\begin{tabularx}{\linewidth}{@{}Xrrrrr@{}}
\toprule
Covariate & All & Convinced & Ambivalent & Disengaged & Rejecters \\
\midrule
Female (\%) & 51.4 & 55.4 & 53.4 & 47.4 & 39.6 \\
Aged 18 to 34 (\%) & 30.5 & 28.6 & 31.3 & 35.0 & 18.1 \\
Aged 55 or over (\%) & 16.6 & 16.5 & 16.6 & 14.4 & 24.2 \\
Household income \pounds{}100,000 or more (\%) & 21.7 & 18.7 & 24.1 & 20.5 & 22.8 \\
EV experience over 3 years (\%) & 24.6 & 20.8 & 27.2 & 22.1 & 31.1 \\
Range below 200 km (\%) & 21.7 & 28.8 & 21.5 & 16.7 & 14.6 \\
Range 400 km or more (\%) & 18.0 & 14.0 & 18.0 & 19.5 & 26.9 \\
No home charging (\%) & 15.2 & 20.4 & 12.5 & 14.2 & 16.1 \\
No off-street parking (\%) & 9.3 & 11.6 & 7.8 & 9.4 & 9.5 \\
Public charging weekly or more often (\%) & 28.3 & 33.5 & 23.0 & 33.7 & 21.5 \\
EV used for commuting (\%) & 78.7 & 82.9 & 77.6 & 78.5 & 70.9 \\
Satisfaction with public charging (mean, 1 to 5) & 3.15 & 3.02 & 3.13 & 3.29 & 3.18 \\
Resident in London (\%) & 10.6 & 9.5 & 9.1 & 16.0 & 5.2 \\
\bottomrule
\end{tabularx}
\end{table}

\begin{table}
\caption{Multinomial logit of class membership, maximum-likelihood three-step estimator: log-odds relative to the ambivalent class (p-value)}\label{tab:mnl}
\small
\begin{tabularx}{\linewidth}{@{}Xrrr@{}}
\toprule
Covariate & Convinced & Disengaged & Rejecters \\
\midrule
Intercept & $-$1.062 (< 0.001) & $-$0.676 (0.026) & $-$1.546 (< 0.001) \\
Female & +0.023 (0.916) & $-$0.335 (0.163) & $-$0.617 (0.053) \\
Age & +0.192 (0.093) & +0.019 (0.881) & +0.381 (0.018) \\
Household income & $-$0.087 (0.420) & $-$0.036 (0.771) & $-$0.233 (0.141) \\
EV experience & $-$0.147 (0.188) & $-$0.142 (0.259) & $-$0.204 (0.215) \\
Vehicle range & $-$0.213 (0.053) & +0.158 (0.198) & +0.300 (0.063) \\
No home charging & +0.485 (0.131) & $-$0.129 (0.736) & +0.708 (0.155) \\
Public-charging dependence & +0.428 (0.120) & +0.557 (0.065) & $-$0.225 (0.619) \\
Satisfaction with public charging & $-$0.077 (0.475) & +0.116 (0.338) & $-$0.017 (0.914) \\
Commuting & +0.460 (0.096) & +0.183 (0.548) & $-$0.048 (0.895) \\
Regional charge points, total & $-$0.002 (0.989) & +0.255 (0.047) & $-$0.148 (0.495) \\
Regional charge points, rapid & $-$0.060 (0.601) & $-$0.021 (0.870) & $-$0.015 (0.932) \\
\bottomrule
\end{tabularx}

\vspace{3pt}
\begin{minipage}{\linewidth}\footnotesize \textit{Note:} The sample contains 728 respondents. The likelihood-ratio test for all covariates is $\chi^2$ = 69.49 on 33 degrees of freedom (p = 0.0002). Continuous covariates are standardised.\end{minipage}
\end{table}

Four patterns are particularly notable.

\begin{itemize}
\item Vehicle range helps distinguish convinced respondents from rejecters. A longer range reduces the odds of being convinced ($-$0.213, p = 0.053) and increases the odds of being a rejecter (+0.300, p = 0.063). Among the convinced group, 28.8\% drive vehicles with a range below 200 km, compared with 14.6\% of rejecters. This echoes the joint-model result linking longer range to lower perceived usefulness (Table~\ref{tab:covlatent}): swapping may be most attractive to drivers for whom range remains a constraint.
\item Rejecters tend to be older and more often male. Age increases the odds of being a rejecter (+0.381, p = 0.018) and, more weakly, of being convinced (+0.192, p = 0.093), suggesting that older respondents appear more often at the ends of the spectrum than in the middle. Being female lowers the odds of rejection ($-$0.617, p = 0.053). Men account for 60.4\% of rejecters, who also tend to have more EV experience and longer-range vehicles.
\item Frequent public charging is common in both the convinced and disengaged groups. About one third in each group uses public charging weekly, compared with around one fifth of the ambivalent and rejecter groups. Dependence on public charging increases the odds of being disengaged (+0.557, p = 0.065), as does regional charge-point density (+0.255, p = 0.047). London residents are over-represented in this group (16.0\%, compared with 10.6\% of the full sample). Where public charging is common and relatively accessible, swapping may seem less necessary; that does not necessarily mean people oppose it.
\item The ambivalent group mostly charges at home. It has the smallest share of respondents without home charging (12.5\%) and one of the lowest shares of weekly public-charging users (23.0\%). Its concerns about swapping therefore largely reflect how people assess the technology, rather than an immediate need to replace home charging.
\end{itemize}

Using an EV for commuting modestly increases the odds of being in the convinced group (+0.460, p = 0.096). Income, EV experience, lack of home charging, satisfaction with public charging and rapid-charger density are not significant predictors of class membership.

\subsection{Summary of hypothesis tests}\label{sec:hyp-summary}

H1 and H3 are supported, while H2 receives partial support. The class analysis also answers the research question by identifying distinct groups of users (Table~\ref{tab:hypotheses}). The table brings together evidence from the construct-level, item-level and class analyses; these results are complementary but do not all test the hypotheses in the same way.

\begin{table}
\caption{Summary of hypothesis tests}\label{tab:hypotheses}
\footnotesize\setlength{\tabcolsep}{4pt}
\begin{tabularx}{\linewidth}{@{}P{0.16}P{0.18}P{0.22}P{0.21}X@{}}
\toprule
Hypothesis & Construct level (joint model) & Item level & Segment level & Assessment \\
\midrule
H1. Perceived usefulness bears more on expected system role than on own purchase preference & Contrast $-$0.178 (SE 0.031), p < 0.001; ratio 0.748. Without RA1: $-$0.190 & Not applicable & Ambivalent class, with higher usefulness, is more confident of the system role than the disengaged class at equal own preference: difference in gaps +0.23, p = 0.016 & Supported; direction robust to correction for single-item measurement error (Section~\ref{sec:structural}) \\
H2. Reasons against bear more on own purchase preference than on expected system role & Contrast $-$0.128 (SE 0.146), p = 0.380; ratio 1.173, in the hypothesised direction & Standardisation $-$0.088 (p = 0.029) and station coverage $-$0.069 (p = 0.081) bear on own preference; operator cost +0.091 (p = 0.019) bears on system role; battery availability $-$0.002 (p = 0.954). Joint Wald $\chi^2$ 12.82, 4 df, p = 0.012 & Consistent: the ambivalent class has the highest barriers and the widest gap & Partially supported: holds for barriers borne by the user, reversed for the barrier borne by the operator \\
H3. The two asymmetries differ & D = 0.306 (SE 0.155), p = 0.049; ratio of ratios 1.568. Without RA1: D = 0.392 (SE 0.307), p = 0.202 & Item contrasts of opposite sign are jointly significant (above) & Not separately testable & Supported in the primary specification; precision depends on the fourth reasons-against indicator \\
RQ. Segments differ in how they combine reasons and in the two outcomes & Not applicable & Not applicable & Four classes (convinced 26.5\%, ambivalent 42.7\%, disengaged 23.1\%, rejecters 7.7\%); 16 of 18 pairwise outcome differences significant; gap between system role and own preference widest in the ambivalent and narrowest in the disengaged class (difference +0.23, p = 0.016) & Answered: four segments that differ on both outcomes, in the pattern H1 and H2 imply \\
Energy-system strain (no directional hypothesis) & Contrast $-$0.070 (SE 0.042), p = 0.095; positive effects on both outcomes & Not applicable & Strain is highest in the convinced and ambivalent classes & Bears more on expected system role, like usefulness \\
\bottomrule
\end{tabularx}
\end{table}

\section{Discussion}\label{sec:discussion}

Among current EV users in the United Kingdom, reasons for and against battery swapping do not simply cancel each other out. They relate differently to personal purchase preference and expectations about the wider system. Perceived usefulness is more closely related to expectations about swapping's future role than to the respondent's own purchase preference. User-facing barriers show the opposite pattern. This contrast is modest when the barriers are combined into one construct, clearer when individual barriers are examined, and also visible across user groups. The discussion below places these findings alongside earlier research and considers what they add to our understanding of battery-swapping acceptance.

\subsection{Perceived usefulness and expectations of the system}\label{sec:disc-usefulness}

Perceived usefulness is the strongest of the three constructs for both outcomes, in line with much of the battery-swapping acceptance literature. Benefit-related beliefs have been linked to adoption intentions among Chinese drivers and members of the public \citep{AduGyamfi2022c,AduGyamfi2022b,XuS2025}, intentions to use swapping in India \citep{Murugan2024}, and continued use among Chinese taxi and ride-hailing drivers \citep{Yan2026}. One Chinese study found that performance expectations were no longer significant after other perceptions were taken into account \citep{Chen2024}. Here, usefulness remains associated with personal purchase preference even after accounting for barriers, perceived energy-system strain and satisfaction with current charging. The result adds evidence from current UK EV users, most of whom charge at home, in a market where private-car swapping is not yet available.

Had the study examined only one outcome, it would not have shown that usefulness is more closely related to expectations about the system than to personal purchase preference. The standardised relationship with personal preference is about three quarters of the relationship with expected system role (ratio 0.748). This differs from \citet{Milani2024}, who found a stronger link between perceived benefits and intention to buy an energy technology than between perceived benefits and intention to support it. The outcomes, however, are not equivalent. Buying, using and supporting a technology are still personal intentions; expected system role asks what someone thinks will happen across the wider market. A respondent can believe that swapping is useful and expect it to matter for transport without wanting to buy a compatible vehicle. Measuring both judgements makes that distinction visible.

The wording of the usefulness items may help explain the difference. Three of the four statements describe swapping as a response to weaknesses in the charging system: insufficient urban infrastructure, limited depot fast charging and slow public charging. Most respondents charge at home and may encounter these problems only occasionally. Their agreement may therefore reflect a view that swapping could help where the existing system falls short, rather than a need they experience on every journey. That kind of assessment is naturally linked to the expected role of swapping in the wider system. The fourth item, which concerns range anxiety, is more personal and has the weakest loading of the four. The findings may thus reflect how this sample understands usefulness as a way to address gaps in charging provision more than a solution to its own daily constraints. Survey items that ask explicitly about personal needs would help distinguish these explanations.

The gap between personal and system-level judgements is consistent with research showing that people may support an innovation for society more than they want it for themselves. People have preferred lower levels of vehicle automation for themselves than for other drivers \citep{Agarwal2024}; support for bike-sharing has exceeded personal intentions to cycle \citep{Nikitas2018}; and individual adoption of energy innovations has been only weakly related to support for energy policies across the European Union \citep{Spandagos2022}. Personal and societal risk judgements can differ as well \citep{TylerCook1984}. Our results add a further point: the same factors do not simply produce different average levels of support; their relationships with the two outcomes also differ in strength. If respondents merely inferred the market's future from their own preferences, each construct would be expected to affect both outcomes in the same proportion. The estimated ratios differ (1.173 compared with 0.748; D is significant at p = 0.049), and this comparison is unaffected by the reliability of the single personal-preference item.

Expectations about a swapping network could also feed back into personal adoption. Previous studies show that expectations about future market shares can shape responses to zero-emission-vehicle policy \citep{Akiyama2025}, that expected sales growth influences choices of swappable electric two-wheelers \citep{Zhang2025a}, and that EV markets face network effects and minimum-scale challenges \citep{Li2017,ZhouLi2018}. Perceived usefulness might therefore encourage personal adoption indirectly by increasing confidence that a viable network will emerge. Our parallel-outcome model compares the two judgements but does not estimate this indirect pathway. Testing it would require repeated observations of the same respondents over time.

\subsection{Reasons against and personal purchase preference}\label{sec:disc-barriers}

The combined reasons-against construct is more closely related to personal purchase preference than to expected system role, although this difference is imprecisely estimated. The item-level results provide clearer evidence. The pattern is in line with BRT studies in which reasons against affect intention directly, while reasons for operate through attitude or other broad evaluations \citep{Claudy2013,Claudy2015,NguyenPhuoc2023,FatahUddin2024}. Expected system role is not an attitude, but it is a general judgement about the technology rather than a decision about one's own behaviour. In both cases, benefits are more closely related to the broader judgement, while obstacles bear more directly on the personal decision.

This distinction may help explain why previous studies disagree about whether positive or negative reasons matter more. Reasons for adoption were stronger in Indian EV studies \citep{Bhat2025,Yadav2024} and research on autonomous vehicles in China \citep{HuangQian2021}. Reasons against were stronger in a study of battery-swapping among Indian two-wheeler users \citep{Chopdar2026} and in research on EV resistance among millennials in Korea and the United States \citep{JuKim2022}. Nearly all these studies measured only one personal outcome. If the balance between positive and negative reasons depends on the outcome being examined, differences across studies may partly reflect how the outcome was defined, as well as genuine differences between populations.

The item-level analysis shows that barriers differ according to who bears their consequences. Battery standardisation was the most widely reported concern (81.7\%) and was more closely related to personal preference (contrast $-$0.088, p = 0.029), with no significant relationship to expected system role. This aligns with findings that interoperability was the leading reason against swapping among Indian two-wheeler users who had used the service \citep{Chopdar2026}, and with engineering reviews that identify incompatible battery formats as a major challenge \citep{Vallera2021,Sen2026}. Our study extends this issue to passenger cars in a market without a swapping service. The concern appears to matter when people consider buying a compatible vehicle, rather than when they judge whether the system will develop. Limited station coverage shows a similar pattern ($-$0.069, p = 0.081). This is consistent with evidence that poor coverage reduces the appeal of swapping even when exchanges are fast \citep{HuangY2021,Zhuge2021}, and that vehicle choice remained low in Japan even when stations were available at many fuel stations \citep{Ito2019}.

Operator inventory cost shows the reverse pattern. It is related to expected system role (contrast +0.091, p = 0.019), but not to personal preference. Respondents may see the cost of holding spare batteries as a question of whether operators can sustain a network, rather than whether swapping would suit them personally. This interpretation is consistent with studies that identify spare-battery inventories and station investment as operator costs \citep{Avci2015,Ito2019} and with the failure of Better Place's swapping business in Denmark and Israel \citep{Noel2016}. It also refines the usual treatment of cost in innovation resistance theory, where cost is often framed as a barrier faced by the adopter \citep{RamSheth1989,AduGyamfi2024b}. A cost borne by the provider may instead create doubt about whether the service will be available at all. Uncertainty about finding a compatible battery (RA4) is not significantly related to either outcome; it may overlap with the standardisation item or may be less clearly understood by respondents.

The partial proportional-odds analysis in Appendix~\ref{app:probit} helps show where barriers matter on the preference scale. Their strongest relationship appears among respondents at the lower end, where barriers distinguish those who would reject a swapping-compatible vehicle from those who might consider one. The barriers therefore seem to provide stronger grounds for rejection than for mild reluctance among people who are already favourable. The survey, however, measured practical barriers only. It did not ask about battery quality or degradation, responsibility for damage, safety or trust in third-party maintenance, all of which feature in other battery-swapping studies \citep{Kegel2025,Geng2025,Yan2026,Liu2026}. Nor did it measure psychological barriers such as tradition and image \citep{RamSheth1989}; tradition was the strongest barrier in one Chinese study of resistance to e-bike swapping \citep{Lu2026}. We cannot determine from these data whether such concerns would relate more strongly to personal preference or to expectations about the system.

\subsection{Perceived energy-system strain}\label{sec:disc-strain}

Respondents who perceive greater pressure on the electricity system are more favourable to battery swapping on both outcomes. As with usefulness, the relationship is somewhat stronger for expected system role (contrast $-$0.070, p = 0.095). Behavioural studies have paid little attention to this connection, although engineering research has examined it in detail. Reviews of battery-swapping operations and grid integration describe how batteries can be charged off-peak, managed around network constraints and used to provide grid services \citep{Revankar2021,Vallera2021,Wu2022,Cui2023,Sen2026}. User-acceptance studies rarely measure how respondents view the electricity system (Section~\ref{sec:lit-acceptance}). The association found here suggests that people who see electrification as a challenge for the grid may view swapping as one way to manage that challenge, rather than as an added burden. This interpretation is consistent with the engineering case for flexible charging.

Two cautions are needed. The survey statements concern electricity supply, renewable energy and network investment in general. Agreement may reflect broader knowledge of energy infrastructure, which could itself make respondents more open to technical solutions; the survey cannot separate this from beliefs about battery swapping. In addition, we measure perceived pressure on the electricity system, not respondents' understanding of how swapping affects electricity demand. The questionnaire did not ask directly whether they thought swapping would increase or reduce network load.

\subsection{Heterogeneity among EV users}\label{sec:disc-heterogeneity}

The four classes reflect distinctions made in the innovation-resistance literature and help explain why users respond differently to swapping. Resistance may mean delaying adoption, rejecting an innovation or actively opposing it \citep{Kleijnen2009}. It can reflect a negative assessment of the innovation itself or satisfaction with the existing option \citep{Heidenreich2015}. Three of the classes fit these patterns, but each has a different combination of benefits, barriers and current charging experience.

\begin{itemize}
\item Rejecters are the clearest example of resistance linked to both a negative view of swapping and satisfaction with current charging. Most are content with their current options (87\%), and no more than 15\% agree with any usefulness statement. They tend to be older, more often male, more experienced EV users and drivers of longer-range vehicles. This resembles \citet{Long2025}, who found that satisfaction with the current vehicle was nearly as important for resistance to zero-emission vehicles as perceived usage benefits. Rejecters form a small share of our sample (7.7\%), broadly consistent with a German study that found a preference for swappable batteries at some additional cost across all respondent groups \citep{Kegel2025}.
\item The ambivalent group shows a different form of resistance. Its members recognise the potential benefits of swapping but report the strongest concerns about all the practical barriers. In the terms of \citet{RamSheth1989}, their hesitation appears to arise from evaluating the technology rather than from contentment with current charging. Despite seeing greater usefulness, their personal purchase preference is similar to that of the disengaged group.
\item The disengaged group is better described as undecided than opposed. Its members choose the middle response nearly three times as often as respondents in the other classes. Their level of disagreement with the outcome statements resembles that of the ambivalent group; the main difference is that they are less likely to agree. Uncertainty is understandable for a service that is not yet available for UK passenger cars. This group also includes more people who use public chargers in areas where charging is relatively accessible, so swapping may seem less necessary to them.
\end{itemize}

The convinced group includes more drivers with shorter-range vehicles, lower satisfaction with current charging and no home charging. Swapping may therefore address constraints that these drivers already experience. This pattern could change over time. If vehicle ranges increase and charging becomes faster and more reliable, the problems that make swapping attractive to this group may become less common. Its current share should not be taken as a fixed estimate of the future market.

The class analysis also shows that the gap between personal and system-level judgements varies across people. The ambivalent group reports high usefulness and high barriers and has the largest gap between its expectations about the system and its own purchase preference. The disengaged group reports less of both and has the smallest gap. Thus, groups with similar personal preferences can still differ in whether they expect swapping to become important. Policies aimed at increasing personal adoption may therefore reach a different audience from measures intended to build confidence in the future network.

\subsection{Methodological implications}\label{sec:disc-methods}

The modelling results offer three lessons for future research on technology acceptance.

First, estimating the measurement and outcome models together changed the interpretation of perceived usefulness. A bivariate ordered probit model using composite scores produced similar usefulness effects for the two outcomes (0.500 and 0.510), whereas the joint model identified a clear difference (Appendix~\ref{app:probit}). Composite scores do not fully account for measurement error, and the resulting estimates may be weakened by different amounts across outcomes. This is one reason to use integrated choice and latent variable models \citep{Daly2012,VijWalker2016}. The two approaches gave similar results for reasons against, but not for usefulness. This matters especially when the research question concerns a difference between effects: if measurement error weakens the estimates unequally, it can hide a meaningful contrast or make one appear where none exists.

Second, latent-variable scaling is not always a purely technical decision. Fixing one loading for each construct while also fixing its disturbance scale imposes an additional restriction. The data reject that restriction (likelihood-ratio statistic 300.44 on four degrees of freedom), and imposing it would distort the scales of constructs whose disturbance variation differs from one. This follows from the model's identification conditions \citep{Daly2012}, but it is easy to overlook in applied work.

Third, combining several barriers into one reflective latent construct can conceal differences that matter substantively. Although all four items describe reasons against swapping, their relationships with the outcomes depend on who bears the consequences. Their different patterns make the overall contrast imprecise. Some BRT studies therefore treat reasons as formative constructs, including the earlier battery-swapping study by \citet{Chopdar2026}. Our findings support examining individual items, or considering a formative specification, when barriers may relate to different outcomes in different ways. Personal preference is also measured with a single item, which limits the correction for measurement error. The comparison is most robust when based on ratios that do not depend on outcome reliability. Future surveys should use several items to measure personal preference, as well as expected system role.

\section{Policy implications and conclusions}\label{sec:policy-conclusions}
This section draws out the policy implications of the findings and then summarises the conclusions, limitations and priorities for future research.
\subsection{Policy implications}\label{sec:policy}

The central policy lesson is that removing barriers for drivers and supporting the viability of a swapping network are related but different goals. Measures that address problems users face are more closely associated with personal adoption, while measures that strengthen the business case are more closely associated with confidence that a network will exist. These goals often involve different organisations. Evaluating one goal using the other outcome could therefore lead to a misleading conclusion about policy effectiveness. Because the study uses cross-sectional survey data from current EV users, the results identify associations rather than causal effects or the size of any policy impact.

\subsubsection{Interoperability standards and personal adoption}\label{sec:policy-standards}

Battery standardisation was the most widely reported concern (81.7\%, and 96\% in the ambivalent group) and had the clearest relationship with personal purchase preference (contrast p = 0.029), with no corresponding relationship to expected system role. Interoperability therefore appears relevant to whether someone would consider buying a vehicle that supports swapping, not only to how the market develops once a network exists. Standardisation is a coordination problem: an individual manufacturer has limited incentive to adopt a common format alone, while a network designed for one brand may serve only that brand's vehicles. Two steps could help. The UK government and standards bodies could take an active role in international standardisation, including the IEC 62840 series on battery swapping, which covers functional, interface and data interoperability \citep{IEC2025}. European standardisation has so far focused mainly on light vehicles \citep{CENCENELEC2025}, while passenger-car swapping in Europe depends on a small number of proprietary systems. International work may therefore be important in establishing a common basis for passenger cars. Public funding for swapping pilots could also require cross-brand compatibility, reducing the risk of locking early investment into proprietary formats. The findings link standardisation concerns to lower personal preference; pilots are still needed to estimate how much common standards would improve acceptance.

\subsubsection{Operator viability and confidence in the network}\label{sec:policy-operators}

The cost of holding charged batteries in reserve is associated with expected system role, but not with personal purchase preference. Financial support for operators, through capital grants, shared-risk arrangements or contracts with fleet operators, may therefore do more to increase confidence that a network can be established than to raise immediate willingness to use it. That confidence still matters. Expectations about future market size can shape responses to vehicle policies \citep{Akiyama2025}, and network effects mean that expected availability can influence EV adoption \citep{Li2017,ZhouLi2018}. Evaluations of operator support should therefore consider changes in confidence and investment, not only immediate changes in stated purchase preference. The failure of Better Place \citep{Noel2016} illustrates why driver interest alone cannot secure a viable business.

\subsubsection{Station coverage and early markets}\label{sec:policy-coverage}

Station coverage, like standardisation, is related to personal purchase preference. To compete with plug-in charging, a swapping network must be practical from a driver's perspective. The results point to possible early markets in places where demand for public charging is high and off-street parking is limited. Respondents who depend on public charging report fewer reasons against swapping, drivers with shorter-range EVs see more usefulness in it, and the convinced group includes more people without home charging and more commuters. Around two in five UK households lack a driveway \citep{DAmico2026}, so swapping could offer an alternative for some households that rely on public charging. Early stations could be tested in locations where public-charging demand is high and private parking is scarce. Commercial fleets, taxis and ride-hailing services have used swapping elsewhere \citep{Yan2026}, but they were not part of this sample and need separate study.

\subsubsection{Grid integration and public communication}\label{sec:policy-grid}

Respondents who perceive greater pressure on the electricity system are more favourable to swapping on both outcomes. They may see it as a way to manage electricity demand. Public information could explain how operators might charge batteries off-peak and adjust charging to local network conditions. Such claims need to be demonstrated in practice and require coordination between swapping operators and electricity distribution network operators. Communication should also acknowledge trade-offs. A swapping network needs spare batteries in addition to those installed in vehicles, which increases demand for battery materials. Its overall environmental benefit will depend on whether smaller vehicle batteries and more intensive battery use compensate for these additional material needs.

\subsubsection{Segment-specific strategies}\label{sec:policy-segments}

The four groups differ in what they need from policy, so a single message or measure is unlikely to work equally well for everyone.

\begin{table}
\caption{Segment-specific measures}\label{tab:segments}
\small
\begin{tabularx}{\linewidth}{@{}lrXX@{}}
\toprule
Segment & Share & Defining profile & Most relevant measures \\
\midrule
Convinced & 26.5\% & High usefulness, discontent with current charging, shorter-range vehicles, more without home charging & Early availability and coverage where they charge; pilots and launch offers aimed at them \\
Ambivalent & 42.7\% & Sees the benefit, endorses every barrier, mostly charges at home & Credible interoperability commitments and assurance that compatible batteries will be stocked; benefit messaging adds little \\
Disengaged & 23.1\% & Withheld judgement, frequent public charging in well-supplied areas & Information and direct experience: demonstrations and trials that resolve uncertainty \\
Rejecters & 7.7\% & Rejects usefulness, content with current charging & No near-term target; barrier removal mainly converts potential rejecters at the margin \\
\bottomrule
\end{tabularx}
\end{table}

Two qualifications should guide any segment-specific strategy in Table~\ref{tab:segments}. The ambivalent group is the largest, but most of its members charge at home; removing barriers may raise their interest without meeting an immediate need. In addition, the partial proportional-odds results in Appendix~\ref{app:probit} suggest that barriers matter most among respondents with low purchase preference. Reducing barriers may therefore move some drivers from rejection to consideration, rather than simply make people who are already favourable even more enthusiastic.

\subsection{Conclusions, limitations and future research}\label{sec:conclusions}

Among current UK EV users, reasons for and against battery swapping relate differently to personal purchase preference and expectations about the wider system. Using an integrated choice and latent variable model, we examined how perceived usefulness, perceived pressure on the electricity system and reasons against swapping relate to both judgements. Usefulness is more closely associated with expected system role than with personal preference, and perceived energy-system strain shows a similar pattern. The overall reasons-against construct points in the opposite direction, although its contrast is imprecisely estimated. The item-level results provide clearer evidence: battery standardisation and station coverage are more closely related to personal preference, whereas the cost of keeping spare batteries is more closely related to expectations about whether a network can operate. The latent class analysis identifies four groups, convinced, ambivalent, disengaged and rejecters. Notably, two groups with similar personal purchase preferences differ in how strongly they expect swapping to shape the future of transport.

These findings extend research on battery-swapping acceptance in four ways. They add evidence from the United Kingdom and from existing EV users, who have received less attention than prospective buyers in China. They also show why it matters to distinguish personal purchase preference from expectations about the wider system: the balance between benefits and barriers depends on the outcome and on who bears the consequences of a barrier. By including perceived pressure on the electricity system, the study connects a concern often discussed in engineering research with users' assessments of swapping. Finally, the four user groups show that the gap between personal and system-level judgements is evident both across explanatory factors and across people.

\subsubsection*{Limitations}

\begin{itemize}
\item Study design. The data are cross-sectional and self-reported. The estimated relationships are associations, not causal effects; the analysis cannot establish whether perceptions shape outcomes or whether people's views of the outcomes shape their perceptions.
\item Sample. Respondents were current EV users recruited through an online panel. They were more affluent and educated than the wider population, and most could charge at home. The results describe this sample rather than all UK drivers, prospective EV buyers or people without off-street parking, who are under-represented.
\item Hypothetical service. Private-car battery swapping is not available in the UK, so respondents assessed a service they had not used. Their views may differ from later behaviour once the service becomes familiar and stations are available. The charging conditions respondents considered in 2025 may also change as charging power and reliability improve.
\item Outcome measurement. Personal purchase preference is measured with a single question, so measurement error may weaken the estimated relationships. Comparing ratios across constructs is not affected by this error. The reasons-against asymmetry is more likely to be understated than exaggerated, and a conservative correction leaves the direction of the usefulness result unchanged (Section~\ref{sec:structural}).
\item Survey coverage. The reasons-against items cover practical barriers but do not measure battery quality or degradation, responsibility for damage, safety, price, or psychological barriers related to tradition and image. In addition, three of the four usefulness questions describe gaps in the charging system rather than constraints that respondents face personally.
\item Context and groups. Regional charger density varies across only eleven regions. The latent classes were identified in one sample, and their labels are interpretations of the response patterns rather than fixed categories of EV users.
\end{itemize}

\subsubsection*{Future research}

\begin{itemize}
\item Future studies should measure personal purchase preference with several items and ask parallel questions about usefulness and barriers. One version could refer to the respondent's own circumstances and another to the wider system. This design would test directly whether the referent explains the differences found here and would allow measurement error to be addressed for both outcomes.
\item Panel studies and experiments could help establish causal effects. For example, researchers could provide different information about interoperability commitments or operator support and then measure whether expectations change. Repeated observations would show whether changes in confidence about the network precede changes in personal purchase preference.
\item Attitudinal measures should also be combined with stated-choice experiments that vary key features of swapping, including price, station coverage, exchange time, battery leasing and responsibility for battery damage. Such experiments could estimate willingness to pay for different service designs \citep{Geng2025,Kegel2025}.
\item The research should be extended to prospective buyers, drivers without off-street parking, commercial fleets and two-wheeler users. The same approach could also be applied to other infrastructure-dependent innovations, such as vehicle-to-grid charging and shared mobility, where personal adoption may differ from expectations about the wider system.
\item Future surveys should vary who bears the consequences of a barrier, the driver or the operator, while keeping the underlying issue constant. This would provide a direct test of whether the distinction identified in the item-level results applies more broadly to the way people evaluate barriers.
\end{itemize}

\appendix

\FloatBarrier
\appsection{A}{Bivariate ordered probit}\label{app:probit}

Before developing the joint model, we estimated a bivariate ordered probit model for personal purchase preference (R1) and expected uptake (R2), allowing the two equations to have correlated errors. This supplementary model used composite scores for the attitudinal constructs. The appendix sets out the specification and estimates, compares them with the joint model, and reports a partial proportional-odds analysis of personal purchase preference.

\subsection{Specification}\label{app:probit-spec}

For each respondent n and outcome h, where h = 1 denotes personal purchase preference and h = 2 denotes expected uptake, the model defines an underlying, unobserved propensity as follows:

\begin{equation}
y_{hn}^{*} = b_{h}' w_{n} + e_{hn}, \qquad \begin{pmatrix} e_{1n}\\ e_{2n}\end{pmatrix} \sim N\left(\begin{pmatrix}0\\ 0\end{pmatrix}, \begin{pmatrix}1 & \varrho\\ \varrho & 1\end{pmatrix}\right)
\label{eq:bop}
\end{equation}

Here, $w_n$ is the vector of predictors, $b_h$ is the coefficient vector for outcome h, and $\varrho$ is the correlation between the two error terms. The observed responses are determined by thresholds applied to each underlying propensity:

\begin{equation}
y_{hn}=m \Leftrightarrow \kappa_{h,m-1} < y_{hn}^{*} \leq \kappa_{hm}, \qquad \kappa_{h0}=-\infty, \quad \kappa_{h5}=+\infty
\label{eq:bopthresholds}
\end{equation}

Each threshold $\kappa_{hm}$ separates one response category from the next and is estimated using the positive-increment approach in Equation~\eqref{eq:thresholds}. The probability of observing a particular pair of responses is:

\begin{equation}
\Pr\left(y_{1n}=m_{1},\,y_{2n}=m_{2}\right) = \Phi_{2}\left(u_{1},u_{2};\varrho\right) - \Phi_{2}\left(l_{1},u_{2};\varrho\right) - \Phi_{2}\left(u_{1},l_{2};\varrho\right) + \Phi_{2}\left(l_{1},l_{2};\varrho\right)
\label{eq:bopprob}
\end{equation}

The limits $u_h$ and $l_h$ define the range of the underlying propensity associated with each observed response category. The bivariate standard normal distribution function $\Phi_2$ accounts for the correlation $\varrho$ between the error terms of the two outcomes.

The predictors include three standardised composite scores, each calculated as the mean of its items: perceived usefulness (PU1--PU4), reasons against (RA2, RA3 and RA4, the three items retained at that stage), and perceived energy-system strain (PES1--PES3). The model also includes satisfaction with current charging and the covariates in Table~\ref{tab:covariates}, except for the regional charger-supply measures: lack of home charging, dependence on public charging, satisfaction with public charging, vehicle range, EV experience, gender, commuting, age and household income. We estimate the model by maximum likelihood, calculate bivariate normal probabilities using 32-point Gauss--Legendre quadrature, and obtain standard errors from the numerical Hessian.

\subsection{Estimates}\label{app:probit-estimates}

\begin{table}
\caption{Bivariate ordered probit of own purchase preference (R1) and expected uptake (R2)}\label{tab:biprobit}
\small
\begin{tabularx}{\linewidth}{@{}Xrrr@{}}
\toprule
Predictor & Own purchase preference: coef. (SE) & Expected uptake: coef. (SE) & Difference: coef. (p) \\
\midrule
Perceived usefulness (composite) & 0.500*** (0.048) & 0.510*** (0.048) & $-$0.010 (0.837) \\
Reasons against (composite) & $-$0.136*** (0.045) & $-$0.015 (0.045) & $-$0.121*** (0.008) \\
Energy-system strain (composite) & 0.096** (0.043) & 0.086** (0.043) & 0.010 (0.811) \\
Status-quo satisfaction & $-$0.197*** (0.048) & $-$0.169*** (0.048) & $-$0.028 (0.569) \\
No home charging & 0.082 (0.124) & $-$0.027 (0.124) & 0.109 (0.376) \\
Public-charging dependence & 0.214** (0.105) & 0.099 (0.105) & 0.115 (0.273) \\
Satisfaction with public charging & 0.000 (0.042) & $-$0.001 (0.043) & 0.001 (0.980) \\
Vehicle range & $-$0.018 (0.041) & 0.012 (0.041) & $-$0.030 (0.465) \\
EV experience & $-$0.045 (0.042) & $-$0.088** (0.043) & 0.043 (0.314) \\
Female & $-$0.072 (0.081) & $-$0.177** (0.081) & 0.104 (0.196) \\
Commuting & 0.228** (0.100) & $-$0.053 (0.101) & 0.281*** (0.005) \\
Age & 0.056 (0.042) & 0.015 (0.043) & 0.041 (0.332) \\
Household income & 0.011 (0.040) & 0.063 (0.041) & $-$0.052 (0.198) \\
Threshold 1 & $-$1.736 (0.128) & $-$2.387 (0.147) &  \\
Threshold 2 & $-$0.757 (0.110) & $-$1.275 (0.115) &  \\
Threshold 3 & 0.375 (0.107) & $-$0.328 (0.107) &  \\
Threshold 4 & 1.494 (0.117) & 0.911 (0.110) &  \\
Error correlation & 0.570*** (0.030) &  &  \\
\bottomrule
\end{tabularx}

\vspace{3pt}
\begin{minipage}{\linewidth}\footnotesize \textit{Note:} N = 728; 35 parameters. Log-likelihood $-$1,734.73; thresholds-only log-likelihood $-$2,082.41; likelihood ratio statistic 695.34 on 27 degrees of freedom (p < 0.001); $\rho^2$ 0.167; AIC 3,539.47; BIC 3,700.13. Wald test that all thirteen coefficients are equal across the two equations: $\chi^2$ 27.79 on 13 degrees of freedom (p = 0.0097). Continuous predictors standardised. *** p < 0.01, ** p < 0.05, * p < 0.10.\end{minipage}
\end{table}

The estimated error correlation is 0.570 (SE 0.030; Table~\ref{tab:biprobit}). This is large enough that estimating the equations separately would discard information and could understate uncertainty. The two outcomes also have some different predictors: the Wald test rejects the hypothesis that all thirteen pairs of coefficients are equal (p = 0.0097). Reasons against are associated with lower personal purchase preference ($-$0.136, p = 0.003), but are not associated with expected uptake ($-$0.015, p = 0.739); the difference between these effects is significant (p = 0.008). Commuting is associated with higher personal preference only, while EV experience and being female are associated with lower expected uptake only.

\subsection{Comparison with the joint model}\label{app:probit-comparison}

The bivariate probit and joint models tell a similar story about reasons against: barriers are more closely related to personal preference than to expectations about the system. Their results differ for perceived usefulness. In the probit model, the coefficients are 0.500 and 0.510 and cannot be distinguished statistically (difference p = 0.837). The joint model instead estimates a standardised contrast of $-$0.178 (p < 0.001). The difference may reflect the use of composite scores in the earlier model. Averaging several Likert-scale items into one score does not account for measurement error, and the resulting estimates may be weakened by different amounts across outcomes. This can conceal a genuine difference between the two relationships.

This supplementary analysis clarifies why the earlier, two-stage estimates for perceived usefulness differ from those of the joint model. It also illustrates a broader risk: a model based on composite scores can suggest that two outcomes have similar predictors even when a joint model using the same data and constructs reveals a difference.

\subsection{Partial proportional-odds analysis}\label{app:ppo}

We also estimated a single-outcome ordered probit model of personal purchase preference to test whether each predictor has the same relationship across the full preference scale. The model includes the three composite scores, satisfaction with current charging, dependence on public charging, commuting, EV experience and gender. For each predictor in turn, we allowed its coefficient to vary across the four response thresholds. We then used a likelihood-ratio test with three degrees of freedom to assess whether one common coefficient was adequate (Table~\ref{tab:ppo}).

\begin{table}
\caption{Partial proportional-odds tests for own purchase preference (ordered probit)}\label{tab:ppo}
\small
\begin{tabularx}{\linewidth}{@{}XQ{0.09}Q{0.09}Q{0.09}Q{0.09}Q{0.09}Q{0.1}Q{0.06}@{}}
\toprule
Predictor & Pooled coefficient & At threshold 1 & At threshold 2 & At threshold 3 & At threshold 4 & LR statistic (3 df) & p \\
\midrule
Perceived usefulness & 0.496 & 0.460 & 0.453 & 0.515 & 0.601 & 2.22 & 0.527 \\
Reasons against & $-$0.135 & $-$0.271 & $-$0.244 & $-$0.121 & $-$0.024 & 8.02 & 0.046 \\
Energy-system strain & 0.101 & 0.042 & 0.127 & 0.039 & 0.196 & 6.79 & 0.079 \\
Status-quo satisfaction & $-$0.204 & $-$0.100 & $-$0.240 & $-$0.203 & $-$0.203 & 2.16 & 0.539 \\
Public-charging dependence & 0.217 & 0.104 & 0.064 & 0.150 & 0.435 & 5.58 & 0.134 \\
Commuting & 0.202 & 0.053 & 0.094 & 0.267 & 0.342 & 2.08 & 0.557 \\
EV experience & $-$0.039 & $-$0.090 & $-$0.016 & 0.003 & $-$0.101 & 3.21 & 0.360 \\
Female & $-$0.073 & $-$0.215 & $-$0.082 & $-$0.047 & $-$0.047 & 0.85 & 0.838 \\
\bottomrule
\end{tabularx}

\vspace{3pt}
\begin{minipage}{\linewidth}\footnotesize \textit{Note:} The sample contains 728 respondents. Threshold 1 separates strong disagreement from all higher response categories; threshold 4 separates strong agreement from all lower categories. Each row allows the effect of one predictor to vary across thresholds.\end{minipage}
\end{table}

Reasons against have their strongest relationship at the lower end of the preference scale: the coefficient is $-$0.271 at the lowest threshold and close to zero at the highest (likelihood-ratio statistic 8.02, p = 0.046). This suggests that barriers mainly distinguish respondents who would reject a swapping-compatible vehicle from those who might consider one. They add little to the distinction among respondents who are already favourable. Energy-system strain shows the reverse pattern, with its largest effect at the top of the scale (p = 0.079). A single coefficient would assume that these constructs affect every point on the preference scale equally, which these results do not support.

\clearpage
\appsection{B}{Applications of Behavioural Reasoning Theory and innovation resistance theory}\label{app:brt}

\noindent
\begin{minipage}{\linewidth}\settabletype
\caption{Applications of Behavioural Reasoning Theory and innovation resistance theory in transport and energy}\label{tab:brt}
\scriptsize\setlength{\tabcolsep}{3pt}\renewcommand{\arraystretch}{1.05}
\begin{tabularx}{\linewidth}{@{}P{0.13}P{0.14}P{0.12}P{0.13}P{0.16}X@{}}
\toprule
Study & Domain, country & Framework & Sample & Outcomes (level) & Relative weight of reasons \\
\midrule
\citet{Claudy2013} & Solar panels, Ireland & BRT & 254 homeowners & Attitude; adoption intention (personal) & Reasons against act directly on intention ($-$0.27); reasons for act only through attitude (total 0.19) \\
\citet{ClaudyPeterson2014} & Commuter cycling, Ireland & BRT & 936 university staff & Attitude; cycling frequency (personal) & Reasons against dominate behaviour ($-$0.50 vs 0.21); attitude has no effect on behaviour \\
\citet{Claudy2015} & Micro wind turbines; car sharing, Ireland & BRT with resistance barriers & 254 homeowners; 379 students and staff & Attitude; adoption intention (personal) & Wind turbines: reasons against dominate ($-$0.31 direct; reasons for act only through attitude). Car sharing: reasons for dominate (0.53 direct; reasons against act only through attitude, $-$0.37) \\
\citet{Chen2021} & E-scooters, Taiwan & BRT & 293 petrol-scooter users & Attitude; switching intention (personal) & Reasons against at least as strong on the direct path to intention; reasons for stronger on attitude \\
\citet{HuangQian2021} & Autonomous vehicles, China & BRT & 849 consumers & Attitude; adoption intention (personal) & Reasons for dominate; reasons against have a small effect on attitude and affect intention only among the risk-averse \\
\citet{JuKim2022} & Electric vehicles, South Korea and US & Innovation resistance & 592 millennial drivers & Resistance (personal) & Perceived risk (0.366) and unavailability (0.335) outweigh relative advantage ($-$0.124) \\
\citet{NguyenPhuoc2023} & Electric motorcycles, Vietnam & BRT & 662 motorcycle users & Attitude, norms, control; switching intention (personal) & Reasons against act only directly on intention; reasons for act only through global motives but have the larger total effect (0.352 vs $-$0.139) \\
\citet{FatahUddin2024} & Electric cars, India & BRT & 312 prospective buyers & Attitude; adoption intention (personal) & Reasons for dominate (0.84 to attitude); reasons against act only on intention ($-$0.11) \\
\citet{Yadav2024} & Electric cars, India & BRT & 338 consumers & Attitude; purchase intention (personal) & Reasons for dominate; reasons against act only through attitude \\
\citet{AduGyamfi2024b} & Battery swapping, cars, China & TAM with innovation resistance & 552 drivers & Adoption intention (personal) & Enablers explain more variance than barriers; usage and value barriers significant, risk not \\
\citet{Bhat2025} & Electric cars, India & BRT with valence framework & 1,243 prospective buyers & Adoption intention, including a recommendation item (personal) & Reasons for dominate (advantages 0.470 vs complexities $-$0.264); perceived fee the strongest reason against \\
\citet{Garibbo2025} & Zero-emission vehicles, Italy & BRT with construal level theory & 1,071 car owners or intending buyers, three groups & Attitude in each of three object conditions (personal) & Reasons for outweigh reasons against for existing product, prospective product and category alike \\
\citet{Long2025} & Zero-emission vehicles, Canada & Innovation resistance with diffusion of innovations & 2,555 prospective buyers & Resistance and purchase intentions; latent class membership (personal) & Satisfaction with the current vehicle about as strong a driver of resistance as perceived usage benefits \\
\citet{Chopdar2026} & Battery swapping, e-two-wheelers, India & BRT & 317 current users & Attitude; use behaviour (personal) & Reasons against dominate use ($-$0.401 vs 0.199) \\
\citet{Lu2026} & Battery swapping, e-bikes, China & Innovation resistance & 503 consumers & Resistance intention (personal) & Tradition barrier strongest; usage and image barriers not significant \\
\bottomrule
\end{tabularx}

\vspace{3pt}
\begin{minipage}{\linewidth}\footnotesize \textit{Note:} BRT: Behavioural Reasoning Theory; TAM: Technology Acceptance Model.\end{minipage}
\end{minipage}

\appsection{C}{Class profiles on the twelve attitudinal indicators}\label{app:class-profiles}

\noindent
\begin{minipage}{\linewidth}\settabletype
\caption{Class profiles on the twelve attitudinal indicators: model-implied mean (\% agree)}\label{tab:classprofiles}
\small
\begin{tabularx}{\linewidth}{@{}Xrrrrr@{}}
\toprule
Indicator & All & Convinced & Ambivalent & Disengaged & Rejecters \\
\midrule
Class share (modal) & 100\% & 26.5\% & 42.7\% & 23.1\% & 7.7\% \\
\addlinespace
\multicolumn{6}{@{}l}{\textit{Perceived usefulness}} \\
PU1, insufficient urban charging & 3.98 (79) & 4.87 (99) & 4.00 (91) & 3.66 (60) & 1.80 (4) \\
PU2, lack of depot fast charging & 3.94 (77) & 4.88 (100) & 3.93 (89) & 3.59 (55) & 1.91 (8) \\
PU3, slow public charging & 3.78 (68) & 4.74 (96) & 3.83 (75) & 3.26 (44) & 1.81 (15) \\
PU4, range anxiety & 3.81 (68) & 4.70 (93) & 3.87 (77) & 3.35 (45) & 1.94 (12) \\
\addlinespace
\multicolumn{6}{@{}l}{\textit{Reasons against}} \\
RA1, too few stations & 3.95 (70) & 4.32 (85) & 4.18 (82) & 3.38 (40) & 3.30 (49) \\
RA2, operator inventory cost & 3.84 (66) & 3.72 (59) & 4.11 (81) & 3.48 (49) & 3.87 (65) \\
RA3, standardisation difficult & 4.20 (82) & 4.14 (81) & 4.51 (96) & 3.62 (57) & 4.50 (88) \\
RA4, unsure of right battery & 3.70 (63) & 3.26 (48) & 4.05 (79) & 3.39 (47) & 4.26 (82) \\
\addlinespace
\multicolumn{6}{@{}l}{\textit{Perceived energy-system strain}} \\
PES1, demand surge & 3.44 (50) & 3.63 (58) & 3.58 (58) & 3.14 (30) & 3.02 (38) \\
PES2, renewables and peak loads & 3.49 (58) & 3.65 (63) & 3.62 (66) & 3.19 (40) & 3.14 (51) \\
PES3, grid inadequate for fleets & 3.89 (73) & 4.12 (79) & 4.04 (84) & 3.40 (50) & 3.78 (71) \\
\addlinespace
\multicolumn{6}{@{}l}{\textit{Status quo}} \\
SAT, happy with current charging & 3.57 (59) & 2.66 (26) & 3.88 (69) & 3.76 (68) & 4.44 (87) \\
Share of all answers at the midpoint & 20.0\% & 14.6\% & 13.4\% & 38.9\% & 13.4\% \\
\bottomrule
\end{tabularx}

\vspace{3pt}
\begin{minipage}{\linewidth}\footnotesize \textit{Note:} Means are calculated from the response probabilities estimated for each class. Percentages in brackets show the share agreeing with each statement (responses 4 or 5).\end{minipage}
\end{minipage}

\FloatBarrier
\bibliographystyle{cas-model2-names}
\bibliography{references}

\end{document}